\documentclass[prd, showpacs,preprintnumbers,nofootinbib,twocolumn]{revtex4-1}
\usepackage[dvips]{graphicx}
\usepackage{enumerate}
\usepackage{amsmath,amssymb}
\usepackage{mathrsfs}
\usepackage{bm}
\usepackage{color}

\begin{document}
\preprint{CHIBA-EP-236, KEK Preprint 2018-82, 2019.09.02}

\title{Type of dual superconductivity for the $SU(2)$ Yang--Mills theory
}

\author{Shogo Nishino${}^{1}$}
\email{shogo.nishino@chiba-u.jp}

\author{Kei-Ichi Kondo${}^{1,2}$}
\email{kondok@faculty.chiba-u.jp}

\author{Akihiro Shibata${}^{3,4}$}
\email{Akihiro.shibata@kek.jp}

\author{Takaaki Sasago${}^{1}$}

\author{Seikou Kato${}^{5}$}
\email{skato@oyama-ct.ac.jp}

\affiliation{
${}^{1}$Department of Physics,
Graduate School of Science, 
Chiba University, Chiba 263-8522, Japan
\\
${}^{2}$Department of Physics,
Graduate School of Science and Engineering, 
Chiba University, Chiba 263-8522, Japan
\\
${}^{3}$Computing Research Center,  High Energy Accelerator Research Organization (KEK), Tsukuba 305-0801, Japan
\\
${}^{4}$SOKENDAI (The Graduate University for Advanced Studies), Tsukuba 305-0801, Japan
\\
${}^{5}$Oyama National College of Technology, Oyama 323-0806, Japan}

\begin{abstract}
We investigate the type of dual superconductivity responsible for quark confinement.
For this purpose, we solve the field equations of the $U(1)$ gauge-scalar model to obtain a single static vortex solution in the whole range without restricting to the long-distance region.
Then we use the resulting magnetic field of the vortex to fit the gauge-invariant chromoelectric field connecting a pair of quark and antiquark which was measured by numerical simulations for $SU(2)$ Yang--Mills theory on a lattice.
This result improves the accuracy of the fitted value for the Ginzburg--Landau parameter to reconfirm the type I dual superconductivity for quark confinement which was claimed by preceding works based on the fitting using the Clem ansatz.
Moreover, we calculate the Maxwell stress tensor to obtain the distribution of the force around the flux tube.
This result suggests that the attractive force acts among chromoelectric flux tubes, in agreement with the type I dual superconductivity.
\end{abstract}

\maketitle

\section{Introduction}

In high energy physics, quark confinement is a long-standing problem to be solved in the framework of quantum field theories, especially quantum chromodynamics (QCD).
The dual superconductivity picture \cite{dsp1,dsp2,dsp3} for the QCD vacuum is known as one of the most promising scenarios for quark confinement.
For a review of the dual superconductivity picture, see, e.g., \cite{PhysRept}.
For this hypothesis to be realized, we must show the existence of some magnetic objects which can cause the dual Meissner effect.
Then, the resulting chromofields are squeezed into the flux tube by the dual Meissner effect.
This situation should be compared with the Abrikosov--Nielsen--Olesen (ANO) vortex \cite{NO1,NO2}  in the $U(1)$ gauge-scalar model as a model describing the superconductor.
In the context of the superconductor, in type I\hspace{-.1em}I the repulsive force acts among the vortices, while in type I the attractive force acts.
The boundary of the type I and type I\hspace{-.1em}I is called the Bogomol'nyi--Prasad--Sommerfield (BPS) limit and no forces act among the vortices.
From the viewpoint of the dual superconductivity picture, the type of dual superconductor characterizes the vacuum of the Yang--Mills theory or QCD for quark confinement.

The type of dual superconductor has been investigated for a long time by fitting the chromoelectric flux obtained by lattice simulations to the magnetic field of the ANO vortex.
The preceding studies \cite{type2-1,type2-2,type2-3,type2-4} done in 1990's concluded that the vacuum of the Yang--Mills theory is of type I\hspace{-.1em}I or the border of type I and type I\hspace{-.1em}I as a dual superconductor.
In these studies, however, the fitting range was restricted to a long-distance region from a flux tube.
The improved studies \cite{Koma1,Koma2} concluded that the vacuum of the Yang--Mills theory can be classified as weakly type I dual superconductor.
Recent studies \cite{Cea-Cosmai1,Cea-Cosmai2,type1-1,type1-2} based on the standard framework of lattice gauge theory, and studies \cite{Kato-Kondo-Shibata, Shibata-Kondo-Kato-Shinohara} based on the new formulation \cite{KSSMKI08,SKS10}, on the other hand, show that the vacua of the $SU(2)$ and $SU(3)$ Yang--Mills theories are strictly type I dual superconductor.
In these works \cite{Cea-Cosmai1,Cea-Cosmai2,type1-1,type1-2,Kato-Kondo-Shibata,Shibata-Kondo-Kato-Shinohara}, the Clem ansatz \cite{Clem} was used to incorporate also the short distance behavior of a flux tube.
The Clem ansatz assumes an analytical form for the behavior of the complex scalar field (as the order parameter of a condensation of the Cooper pairs), which means that it still uses an approximation.
In this work, we shall fit the chromoelectric flux tube to the magnetic field of the ANO vortex in the $U(1)$ gauge-scalar model without any approximations to examine the type of dual superconductor.
Indeed, we determine the Ginzburg--Landau (GL) parameter by fitting the lattice data of the chromoelectric flux to the numerical solution of the ANO vortex in the whole range.
The resulting value of the GL parameter reconfirms that the dual superconductivity of $SU(2)$ Yang--Mills theory is of type I.

In addition, in order to estimate the force acting among the flux tubes, we investigate the Maxwell stress force carried by a single vortex configuration.
Recently, the Maxwell stress force distribution around a quark-antiquark pair was directly measured on a lattice via the gradient flow method \cite{EMT}.
Our results should be compared with theirs.
For this purpose, we shall calculate the energy-momentum tensor originating from a single ANO vortex solution to obtain the distribution of the Maxwell stress force corresponding to the obtained value of the GL parameter.

This paper is organized as follows.
In Section $\rm{I\hspace{-.1em}I}$, we introduce an operator to measure chromofields produced by a pair of quark and antiquark on a lattice.
We review the results of lattice measurements in \cite{Kato-Kondo-Shibata}.
In Section $\rm{I\hspace{-.1em}I\hspace{-.1em}I}$, we give a brief review of the ANO vortex in the $U(1)$ gauge-scalar model.
Then, we discuss the type of superconductor characterized by the GL parameter.
In Section $\rm{I\hspace{-.1em}V}$, we explain a new method of fitting after giving a brief review of the fitting method based on the Clem ansatz adopted in the previous study \cite{Kato-Kondo-Shibata} in order to compare our new result with the previous one. 
In Section V, we study the distribution of the force around a single flux tube by considering the Maxwell stress tensor.
In Section $\rm{V\hspace{-.1em}I}$, we summarize our results.
In Appendix A, we explain the advantage of the operator which we propose based on the new formulation to measure the gauge-invariant field strength on a lattice.

\section{Operator on a lattice to measure the flux tube}

\begin{figure}[t]
\centering
\includegraphics[width=0.3\textwidth]{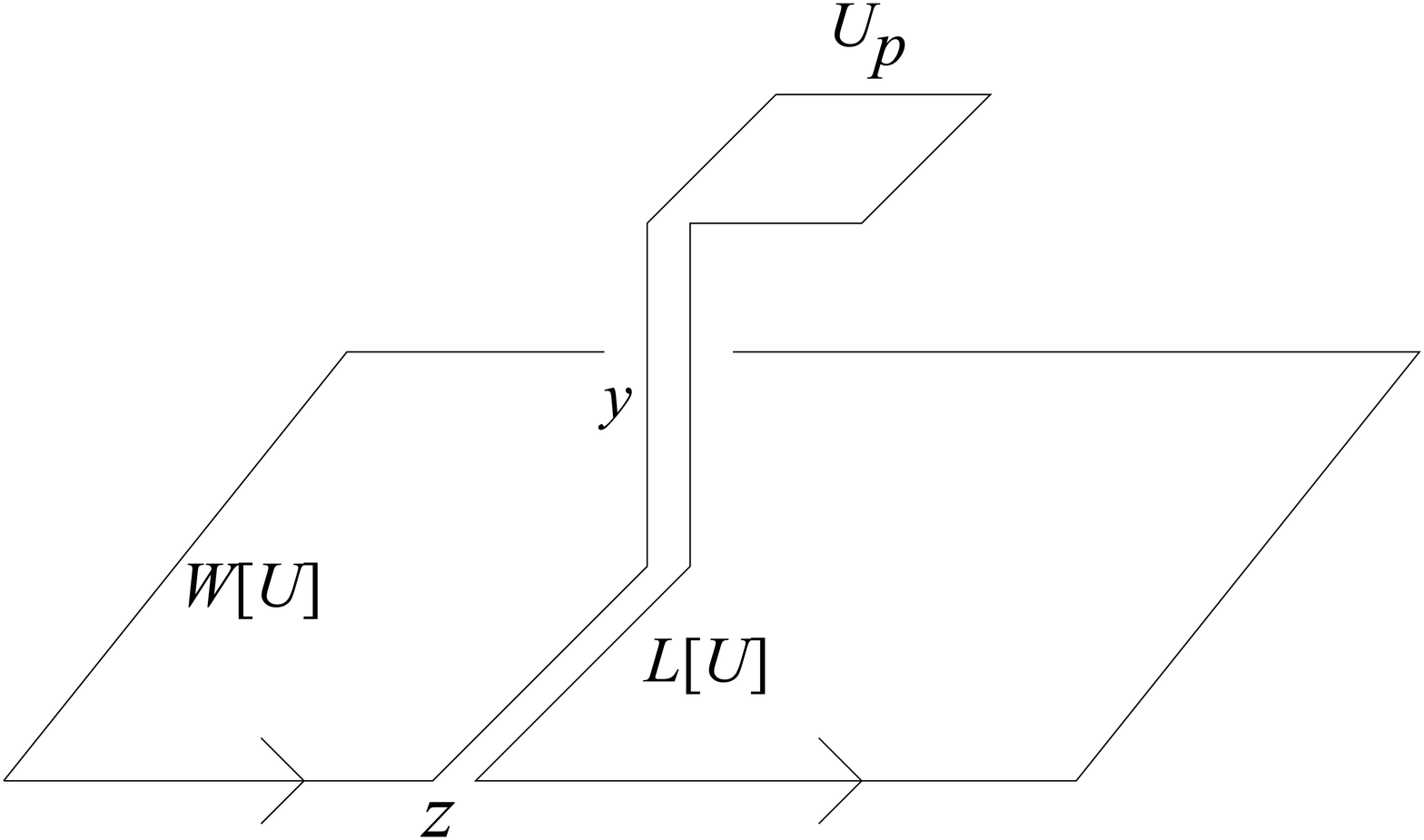} 
\caption{The setup of the operator $W [U] L [U] U_{P} L^{\dagger} [U]$ in (\ref{lattice_operator_U}).
$z$ is the position at which the Schwinger line $L [U]$ is inserted, and $y$ is the distance from the Wilson loop $W [U]$ to the plaquette $U_{P}$.
}
\label{lattice_result1}
\end{figure}

\begin{figure*}[t]
\centering
\includegraphics[width=0.4\textwidth]{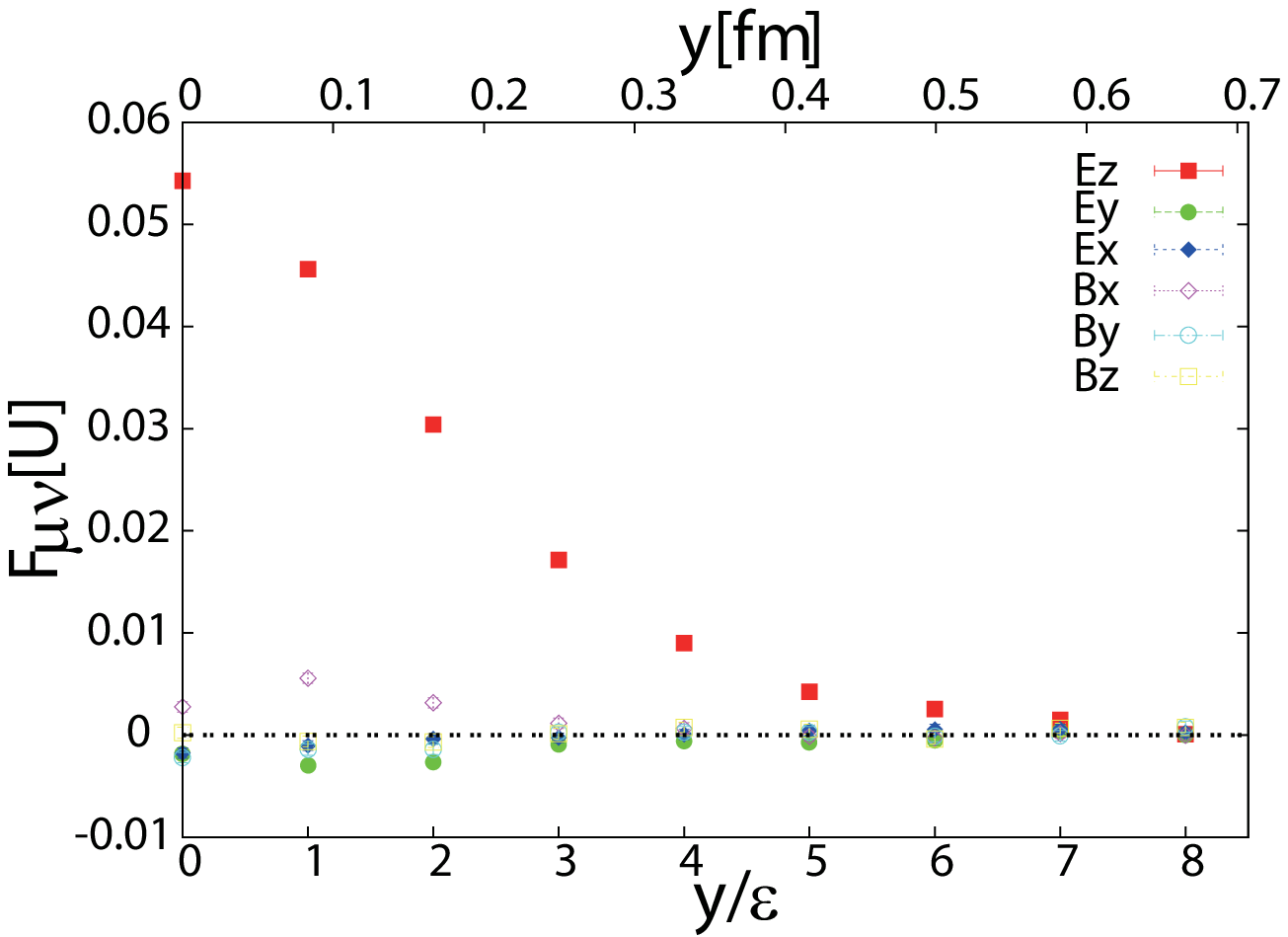} \ \ \ 
\includegraphics[width=0.45\textwidth]{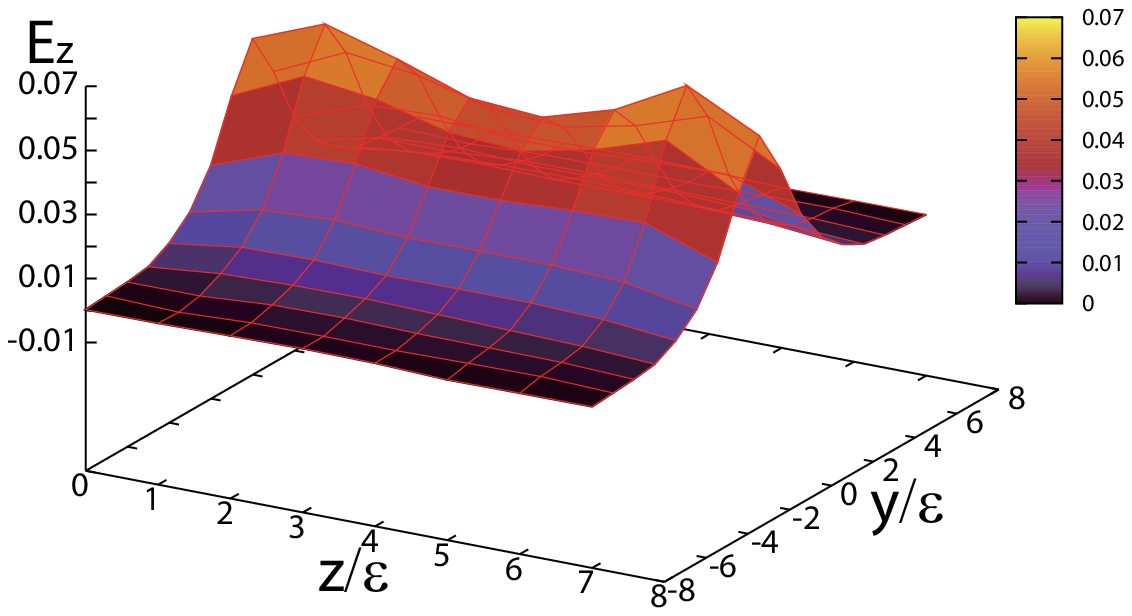}
\caption{ \cite{Kato-Kondo-Shibata} 
(Left panel) The gauge-invariant chromofields $F_{\mu \nu} [U]$ in (\ref{strength_U}) at the midpoint of the $q \bar{q}$ pair ($z =4$) for the $8 \times 8$ Wilson loop  on the $24^{4}$ lattice with the lattice spacing $\epsilon = 0.08320 \ \mathrm{fm}$ at $\beta = 2.5$ .
(Right panel) The distribution of $E_{z} [U] = F_{34} [U]$ in $y-z$ plane.}
\label{lattice_result}
\end{figure*}

\begin{figure*}[t]
\centering
\includegraphics[width=0.4\textwidth]{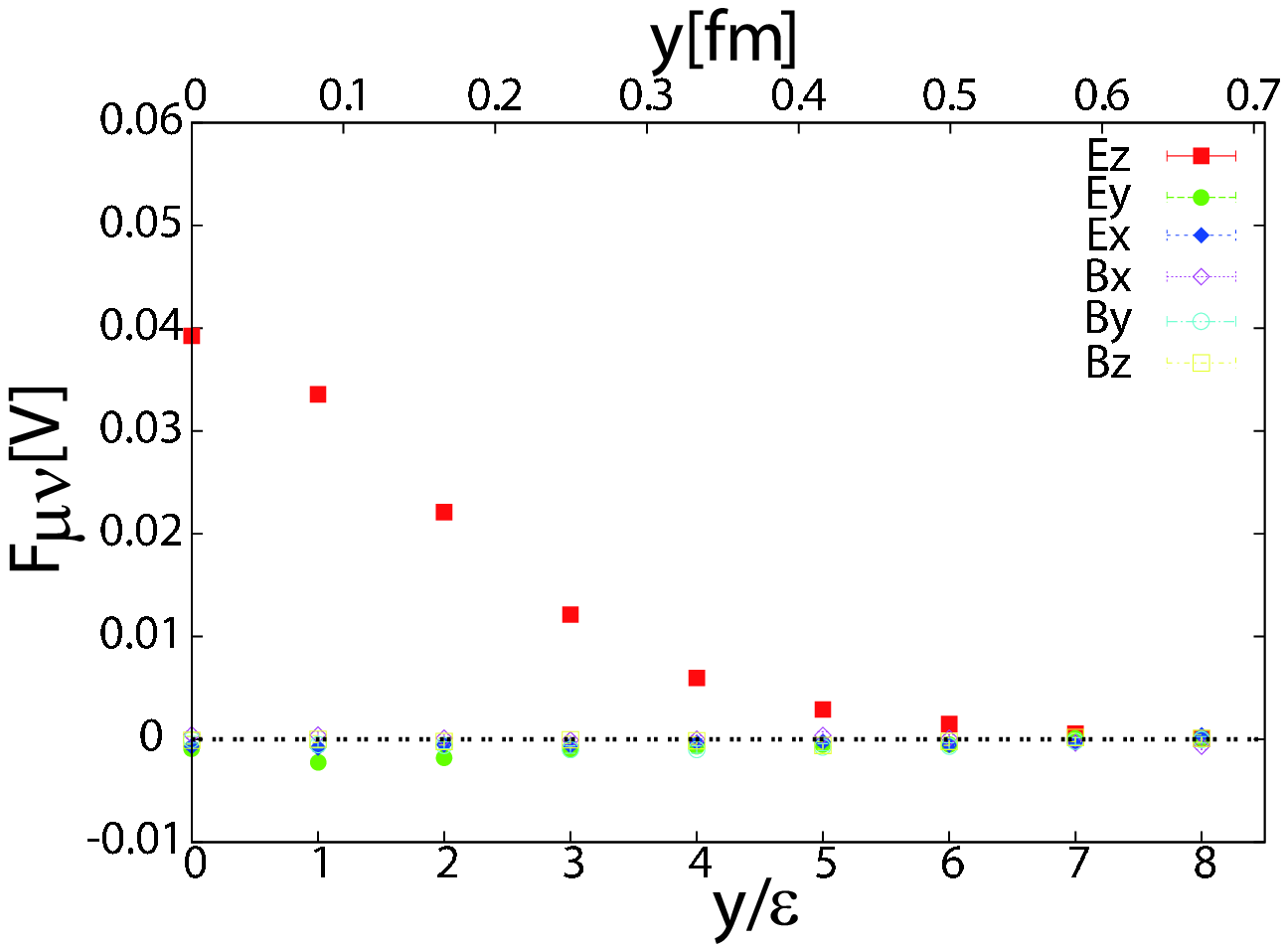} \ \ \ \ \ \ 
\includegraphics[width=0.45\textwidth]{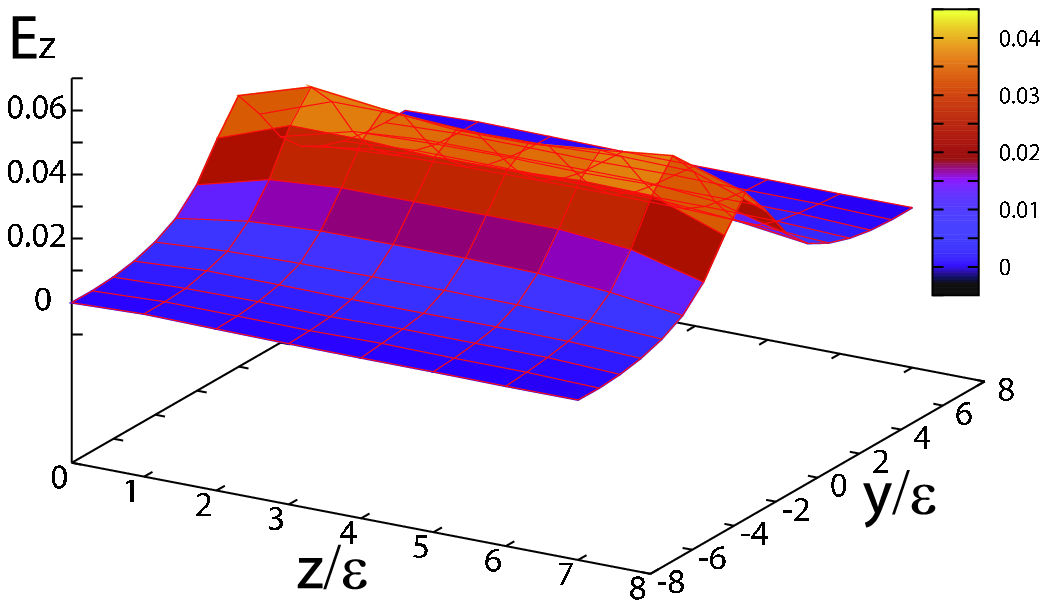} 
\caption{\cite{Kato-Kondo-Shibata} 
(Left panel) The gauge-invariant chromofields $F_{\mu \nu} [V]$ in (\ref{strength_V}) under the same conditions as those in FIG.\ref{lattice_result}.
(Right panel) The distribution of $E_{z} [V]$ in $y-z$ plane.}
\label{lattice_result_V}
\end{figure*}

In order to measure the chromofield strength $\mathscr{F}_{\mu \nu}$ generated by a pair of  static quark and antiquark belonging to the fundamental representation of the gauge group $G = SU (2)$, we use the gauge-invariant operator proposed by Di Giacomo, Maggiore, and Olejnik \cite{Giacomo} using the Wilson loop operator $W [U]$ defined by the Yang--Mills link variable $U \in SU (2)$ along a path $C$ ($L \times T$ rectangular):\footnote{We use the notation $\rho [U]$ to indicate the average coming from the operators defined in terms of the original link variable $U$, since we define the similar operator defined in terms of the different variable later.}
\begin{align}
\rho [U] := & \frac{\left\langle \mathrm{tr} \left( W [U] L[U] U_{P} L^{\dagger} [U] \right) \right\rangle}{\langle \mathrm{tr} ( W [U] ) \rangle} 
\nonumber\\
&\hspace{1cm} - \frac{1}{\mathrm{tr}( \bm{1})} \frac{\langle \mathrm{tr} ( U_{P} ) \mathrm{tr} ( W [U] ) \rangle}{\langle \mathrm{tr} ( W [U] ) \rangle}
,
\label{lattice_operator_U}
\end{align}
where $U_{P}$ is a single plaquette constructed by $U$ and $L [U]$ is called the Schwinger line operator connecting the Wilson loop operator $W [U]$ and the plaquette $U_{P}$.
See FIG.\ref{lattice_result1} for the setup of the operator $W[U] L[ U] U_{P} L^{\dagger} [U]$.
In the continuum limit where the lattice spacing $\epsilon$ vanishes $\epsilon \to 0$, $\rho [U]$ reduces  to
\begin{align}
\rho [U] = & i g \epsilon^{2} \frac{\left\langle \mathrm{tr} ( \mathscr{F}_{\mu \nu} [\mathscr{A}] L^{\dagger} [U] W [U] L [U] ) \right\rangle}{\langle \mathrm{tr} ( W [U] ) \rangle} + \mathcal{O} (\epsilon^{4} )  \nonumber\\
\simeq & g \epsilon^{2} \langle \mathscr{F}_{\mu \nu} [\mathscr{A}] \rangle_{q \bar{q}}
,
\end{align}
where $\mathscr{A} \in \mathfrak{su} (2)$ stands for the gauge field of the continuum $SU(2)$ Yang--Mills theory, which is related to the link variable $U$ as $U_{x, \mu} = \exp \left(- i g \epsilon \mathscr{A}_{\mu} (x) \right)$. 
Thus, the field strength $F_{\mu \nu} [U]$ generated by a quark-antiquark pair can be obtained by
\begin{equation}
F_{\mu \nu} [U] = \frac{\sqrt{\beta}}{2} \rho [U] , \ \ \ \beta = \frac{4}{g^{2}}
.
\label{strength_U}
\end{equation}
FIG.\ref{lattice_result} shows the chromofield strength $F_{\mu \nu} [U]$ measured at the midpoint of the $q \bar{q}$ pair for the $8 \times 8$ Wilson loop on the $24^{4}$ lattice at $\beta = 2.5$ \cite{Kato-Kondo-Shibata}.
In this paper, we have fixed the physical scale of the lattice spacing $\epsilon = 0.08320 \ \mathrm{fm}$ at $\beta = 2.5$ for $SU(2)$ by fixing the physical string tension $\sigma_{\rm phys} = (440 \ \mathrm{MeV})^{2}$ according to the relation $\sigma_{\rm lat} = \sigma_{\rm phys} \epsilon^{2}$ \cite{Kato-Kondo-Shibata}.
Our results are consistent with the preceding studies \cite{Cea-Cosmai1,Cea-Cosmai2}.

In the previous study \cite{Kato-Kondo-Shibata}, we used the new formulation \cite{KSSMKI08,SKS10} of the lattice Yang--Mills theory by decomposing the gauge field $U_{x,\mu}$ into $V_{x,\mu}$ and $X_{x,\mu}$, $U_{x,\mu} = X_{x,\mu} V_{x,\mu}$, where $V_{x,\mu} \in SU (2)$ called the {\it restricted link variable} is supposed to have the same transformation law as the original link variable $U$ under the gauge transformation, and a remaining part $X_{x,\mu} \in SU(2)$ called the {\it remaining site variable} transforms in an adjoint way under the gauge transformation.
The restricted link variable $V_{x,\mu}$ plays a very important role for realizing the dual superconductor picture, since the dominant mode for quark confinement is extracted from it, for example, $V_{x,\mu}$ induces naturally the magnetic current.
See, e.g., \cite{PhysRept} for more details.

In the new formulation, the key ingredient is the color direction field  $\bm{n}_{x}$ which takes the value in the $SU(2)$ Lie algebra with a constraint of a unit length 
$\bm{n}_{x} \cdot \bm{n}_{x}=1$ and transforms in an adjoint way under the gauge transformation.
The color direction field $\bm{n}_{x}$ is in advance obtained as a functional of the original link variable $U_{x,\mu}$ by solving the reduction condition \cite{IKKMSS06,PhysRept}. 
Then the restricted link variable $V_{x,\mu}$ is expressed in terms of the original link variable $U_{x,\mu}$  and  the color direction field $\bm{n}_{x}$ as   
\begin{align} 
V_{x,\mu} := 
 \tilde{V}_{x,\mu}/\sqrt{\frac{1}{2}{\rm tr} [\tilde{V}_{x,\mu}^{\dagger}\tilde{V}_{x,\mu}]} ,
\nonumber\\
  \tilde{V}_{x,\mu} 
  := U_{x,\mu} +   \bm{n}_{x} U_{x,\mu} \bm{n}_{x+\mu} .
  \label{sol}
\end{align}
See Appendix A for more details.

In view of these, we propose to use the operator $\rho [V]$ similar to (\ref{lattice_operator_U}) by replacing the full link variable $U$ by the restricted link variable $V$:
\begin{align}
\rho [V] := & \frac{\left\langle \mathrm{tr} \left( W [V] L [V] V_{P} L^{\dagger} [V] \right) \right\rangle}{\langle \mathrm{tr} ( W [V] ) \rangle} \nonumber\\
&\hspace{1cm} - \frac{1}{\mathrm{tr} (\bm{1})} \frac{\langle \mathrm{tr} ( V_{P} ) \mathrm{tr} ( W [V] ) \rangle}{\langle \mathrm{tr} ( W [V] ) \rangle}
,
\label{lattice_operator_V}
\end{align}
where $W [V]$ is the restricted Wilson loop operator obtained by replacing the link variable $U$ by $V$.
In the continuum limit $\epsilon \to 0$, $\rho [V]$ reduces to
\begin{align}
\rho [V] \simeq & g \epsilon^{2} \langle \mathscr{F}_{\mu \nu} [\mathscr{V}] \rangle_{q \bar{q}} , 
\label{continuum_limit}
\end{align}
and therefore, we can define the chromofield strength $F_{\mu \nu} [V]$  generated by $q \bar{q}$ pair  for the restricted link variable $V$ by
\begin{align}
F_{\mu \nu} [V] = & \frac{\sqrt{\beta}}{2} \rho [V] , \ \ \ \beta  = \frac{4}{g^{2}} .
\label{strength_V}
\end{align}
FIG.\ref{lattice_result_V} shows the restricted chromofield strength $F_{\mu \nu} [V]$ measured in the same settings as $F_{\mu \nu} [U]$ \cite{Kato-Kondo-Shibata}.

In Appendix A, we demonstrate advantages of using $\rho [V]$ constructed from the restricted link variable $V$ based on the new formulation, in sharp contrast to the preceding operator $\rho [U]$ defined in terms of the original link variable $U$ based on the ordinary framework of lattice gauge theory:  
(i) The operator $\rho [V]$ enables us to extract the non-trivial gauge-invariant and Abelian-like field strength which is used to measure the chromoelectrix flux, in sharp contrast to the gauge-covariant non-Abelian field strength. 
(ii) The operator $\rho [V]$ does not depend on the choice of the Schwinger lines $L, L^\dagger$, namely, the shape of $L, L^\dagger$ and the position $z$ at which the Schwinger lines are inserted. 

\begin{figure*}[t]
\centering
\includegraphics[width=0.25\textwidth]{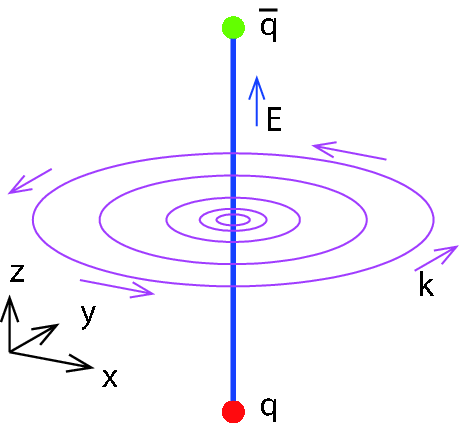} \ \ \ 
\includegraphics[width=0.45\textwidth]{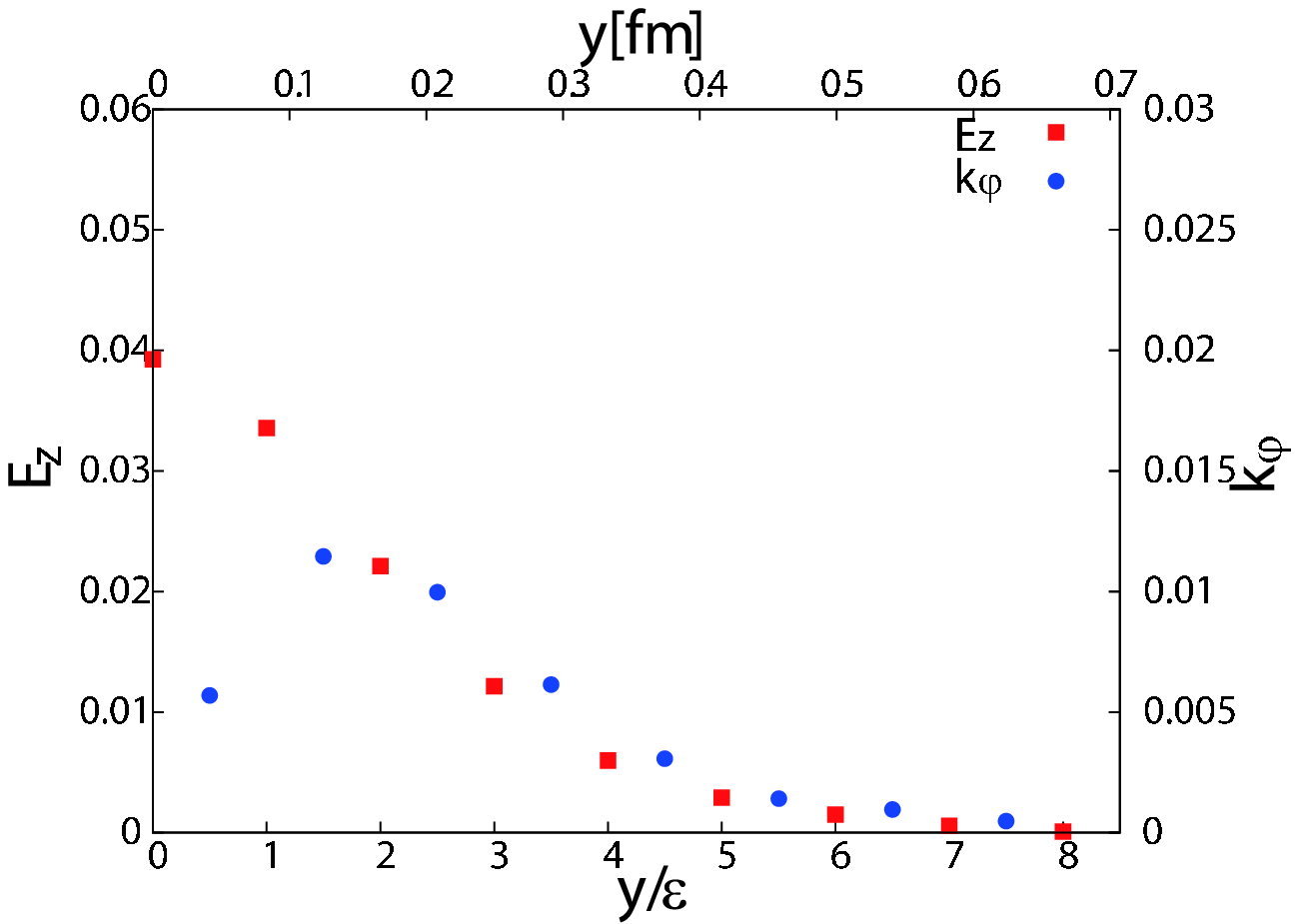}
\caption{ \cite{Kato-Kondo-Shibata} 
(Left panel) The relation among the chromoelectric field $\bm{E}$, the induced magnetic current $k_{\mu}$, and the quark-antiquark pair $q \bar{q}$. 
(Right panel) The induced magnetic current $k_{\mu}$ obtained by (\ref{current_lattice}) using the chromofield $F_{\mu \nu} [V]$ in (\ref{strength_V}).}
\label{lattice_current}
\end{figure*}

Recent study \cite{renormalize} suggests that the operator $\rho [U]$ undergoes nontrivial renormalizations, which depend on the length and on the number of cusps in the Schwinger lines.
The study \cite{Baker} suggests that the extended smearing behaves like an effective renormalization of the operator $\rho [U]$.
For $\rho [V]$, however, such renormalizations are not necessary since $\rho [V]$ does not depend on the Schwinger lines.
On the other hand, renormalization or smearing for the restricted Wilson loop operator and the probe should be taken into account.
In the previous study \cite{Kato-Kondo-Shibata}, we used the hypercubic blocking (HYP) method \cite{HYP} once for the link variables on the Wilson loop to reduce high-energy noises for both $U$ and $V$.
However, we find numerically that for the restricted field $V$, the measured expectation value hardly differs from the unsmeared case.

First of all, we observe that the $z$-component of the restricted chromoelectric field $E_{z} [V]$ forms a uniform flux tube compared with a non-uniform one $E_{z} [U]$ \cite{Kato-Kondo-Shibata, Shibata-Kondo-Kato-Shinohara}, since the effect due to the static sources placed at a finite distance in $E_{z} [V]$ is smaller than $E_{z} [U]$.
Therefore, the restricted chromoelectric flux $E_{z} [V]$ can be well approximated by the ANO vortex with an infinite length.
Moreover, it was shown in the previous studies \cite{Kato-Kondo-Shibata, Shibata-Kondo-Kato-Shinohara} that the type of dual superconductor determined only by the flux tube does not change irrespective of whether we use $E_{z} [U]$ or $E_{z} [V]$.
By these reasons, we shall use the data of $E_{z} [V]$ for fitting.

It should be noticed that we can define the magnetic current $k_{\mu}$ induced by the chromofield $F_{\mu \nu} [V]$ as
\begin{equation}
k_{\mu} := \frac{1}{2} \epsilon_{\mu \nu \rho \sigma} \nabla_{\nu} F_{ \rho \sigma} [V]
,
\label{current_lattice}
\end{equation}
with the lattice derivative $\nabla_{\nu}$ so that the conservation law $\nabla_{\mu} k_{ \mu} = 0$ holds  \cite{Kato-Kondo-Shibata, Shibata-Kondo-Kato-Shinohara}.
Since the nontrivial component of the chromofield $F_{\mu \nu} [V]$ is only the $z$-component $E_{z} [V]$ of the chromoelectric field (see the left panel of FIG.\ref{lattice_result_V}), the induced magnetic current $k_{\mu}$ has only the component $k_{\varphi}$ circulating around a flux tube.
The left panel of FIG.\ref{lattice_current} is an illustration of the relation between the chromoelectric field $\bm{E}$ and the induced magnetic current $k_{\mu}$.
The right panel of FIG.\ref{lattice_current} is a plot of the chromoelectric field $E_{z} [V]$ and the magnetic current $k_{\varphi}$ induced around a single chromoelectric flux tube.

\section{The gauge-scalar model and type of superconductor}
\subsection{The Abrikosov--Nielsen--Olesen vortex}

In this subsection, we give a brief review of the $U(1)$ gauge-scalar model with the Lagrangian density given by
\begin{equation}
\mathscr{L} = - \frac{1}{4} F_{\mu \nu} F^{\mu \nu} + \left( D_{\mu} \phi \right)^{\ast} D^{\mu} \phi - \frac{\lambda^{2}}{2} \left( \phi^{\ast} \phi - v^{2} \right)^{2}
,
\label{L_AH}
\end{equation}
where $\lambda$ is a coupling constant of the scalar self-interaction,  and $v$ is a value of the magnitude $|\phi (x)|$ of the complex scalar field $\phi (x)$ at the vacuum $|x| = \infty$.
The asterisk $({}^{\ast})$ denotes the complex conjugation.
The field strength $F_{\mu \nu}$ of the $U(1)$ gauge field $A_{\mu}$ and the covariant derivative $D_{\mu} \phi$ of the scalar field $\phi$ are defined by
\begin{align}
F_{\mu \nu} (x) := & \partial_{\mu} A_{\nu} (x) - \partial_{\nu} A_{\mu} (x) , \\
D_{\mu} \phi (x)  := & \partial_{\mu} \phi (x) - i q A_{\mu} (x) \phi (x)  ,
\end{align}
where $q$ is the electric charge of the scalar field $\phi (x)$.
The Euler--Lagrange equations are given as
\begin{align}
D^{\mu} D_{\mu} \phi = & \lambda^{2} \left( v^{2} - \phi^{\ast} \phi \right) \phi , \label{eq_scalar}\\
\partial^{\mu} F_{\mu \nu} = & j_{\nu} ,\label{eq_gauge}
\end{align}
where we define the electric current $j_{\mu}$ by
\begin{align}
j_{\nu} := &  i q \bigl[ \phi \left( D_{\nu} \phi \right)^{\ast} - \left( D_{\nu} \phi \right) \phi^{\ast} \bigr] 
.
\end{align}

In order to describe the vortex solution, we introduce the cylindrical coordinate system $(\rho , \varphi , z)$ for the spatial coordinates with the associated unit vectors $\bm{e}_{\rho}, \bm{e}_{\varphi},$ and $\bm{e}_{z}$, and adopt a static and axisymmetric ansatz:
\begin{equation}
A_{0} (x) = 0 , \ \ \ 
\bm{A} (x) = A (\rho) \bm{e}_{\varphi} , \ \ \ 
\phi (x) = v f (\rho) e^{i n \varphi}
,
\label{ANO_ansatz}
\end{equation}
where $n$ is an integer.
Under this ansatz, the field equations (\ref{eq_gauge}) and (\ref{eq_scalar}) are cast into
\begin{align}
& - \frac{1}{\rho} \frac{d}{d \rho} \biggl[ \rho \frac{d}{d \rho} f (\rho) \biggr] + \biggl[ \frac{n}{\rho} - q A (\rho) \biggr]^{2} f (\rho) \nonumber\\
&\hspace{2cm}= \lambda^{2} v^{2} \bigl[ 1 - f^{2} (\rho) \bigr] f (\rho) , \label{pre_eq_f}\\
& \frac{d}{d \rho} \biggl[ \frac{1}{\rho} \frac{d}{d \rho} \left( \rho A (\rho ) \right) \biggr] = j_{\varphi} (\rho) ,
\label{pre_eq_A}
\end{align}
where a non-vanishing component $j_{\varphi}$ of  the electric current is written as
\begin{align}
&j_{\varphi} (\rho) =  2 q^{2} v^{2} \biggl[ A (\rho) - \frac{n}{q \rho} \biggr] f^{2} (\rho)
\label{pre_eq_j}
.
\end{align}
Moreover, the magnetic field $\bm{B}$ is given in the present ansatz by
\begin{equation}
\bm{B} (x) =  \nabla \times \bm{A} (x) = \frac{1}{\rho} \frac{d}{d \rho} \left( \rho A (\rho) \right) \bm{e}_{z}
.
\label{B_ANO}
\end{equation}

To determine the boundary conditions, let us consider the static energy $E$.
The energy-momentum tensor $T^{\mu \nu}$ is obtained from the Lagrangian density (\ref{L_AH}) as
\begin{align}
T^{\mu \nu} = & \frac{1}{4} g^{\mu \nu} F_{\rho \sigma} F^{\rho \sigma} - F^{\mu \rho} F^{\nu}{}_{\rho} + \left( D^{\mu} \phi \right) \left( D^{\nu} \phi \right)^{\ast}  \nonumber\\
&+ \left( D^{\mu} \phi \right)^{\ast} \left( D^{\nu} \phi \right) - g^{\mu \nu} \left( D_{\rho} \phi \right) \left( D^{\rho} \phi \right)^{\ast} \nonumber\\
& + \frac{\lambda^{2}}{2} g^{\mu \nu} \left( v^{2} - \phi^{\ast} \phi \right)^{2}
.
\label{EMT_ANO}
\end{align}
Notice that this energy-momentum tensor is symmetric, i.e., $T^{\mu \nu} = T^{\nu \mu}$.
Then, the static energy $E$ is obtained as
\begin{align}
E = & \int d^{3} x \ T^{0 0} \nonumber\\
= & 2 \pi \int_{- \infty}^{\infty} d z \int_{0}^{\infty} d \rho \ \rho \left\{ \frac{1}{2} \frac{1}{\rho^{2}} \biggl[ \frac{d}{d \rho} \left( \rho A (\rho) \right) \biggr]^{2} \right. \nonumber\\
&+ v^{2} \biggl[ \frac{d}{d \rho} f (\rho) \biggr]^{2} 
 +  v^{2} \biggl[ \frac{n}{\rho} - q A (\rho) \biggr]^{2} f^{2} (\rho) \nonumber\\
&+\left. \frac{\lambda^{2} v^{4}}{2} \bigl[ 1 - f^{2} (\rho) \bigr]^{2} \right\}
.
\label{energy_ANO}
\end{align}
In what follows, we consider the energy per unit length of a vortex to avoid the divergence, since the energy density $T^{00}$ does not depend on $z$.

The static energy $E$ given by (\ref{energy_ANO}) is nonnegative, $E \geq 0$.
The equality $E=0$ holds if and only if
\begin{equation}
f (\rho) = 1 , \ \ \ A (\rho) = \frac{n}{q \rho}
,
\label{vacuum_sol}
\end{equation}
are satisfied.
Since the equation (\ref{vacuum_sol}) is the solution of the field equations (\ref{pre_eq_f}) and (\ref{pre_eq_A}), we call it the vacuum solution.

\begin{figure*}[t]
\centering
\includegraphics[width=0.4\textwidth]{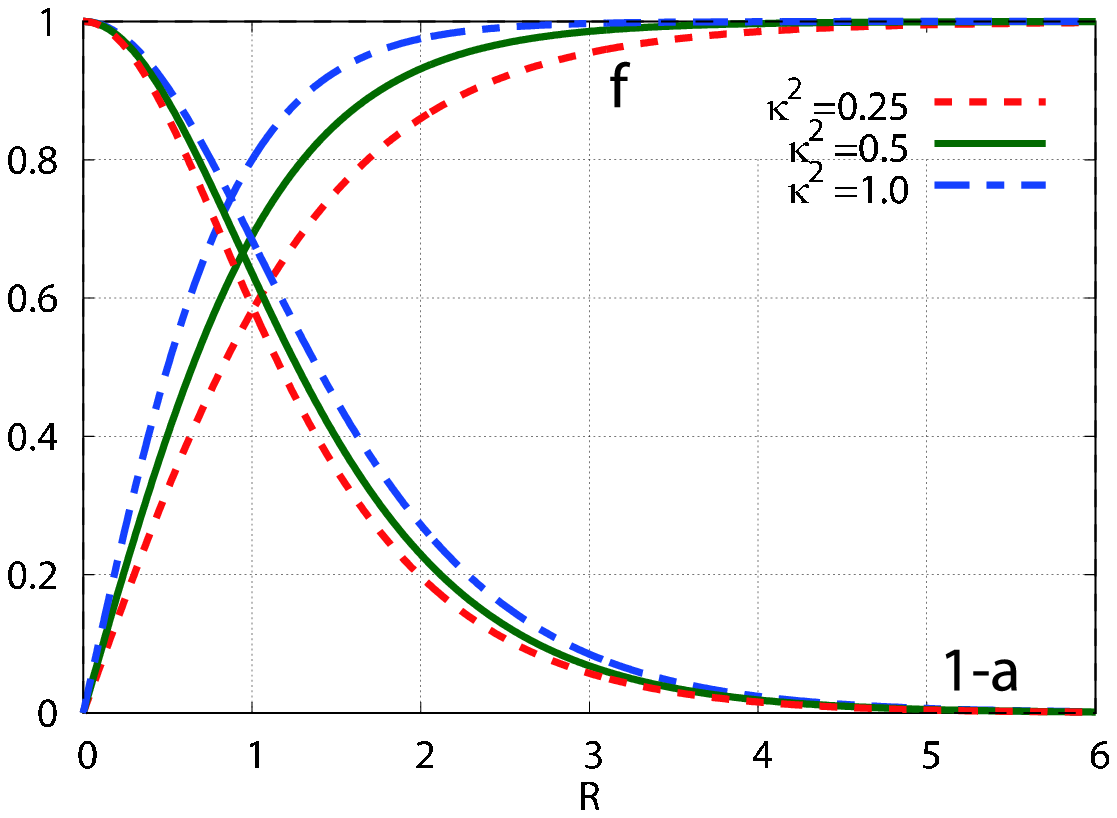} \ \ \ 
\includegraphics[width=0.4\textwidth]{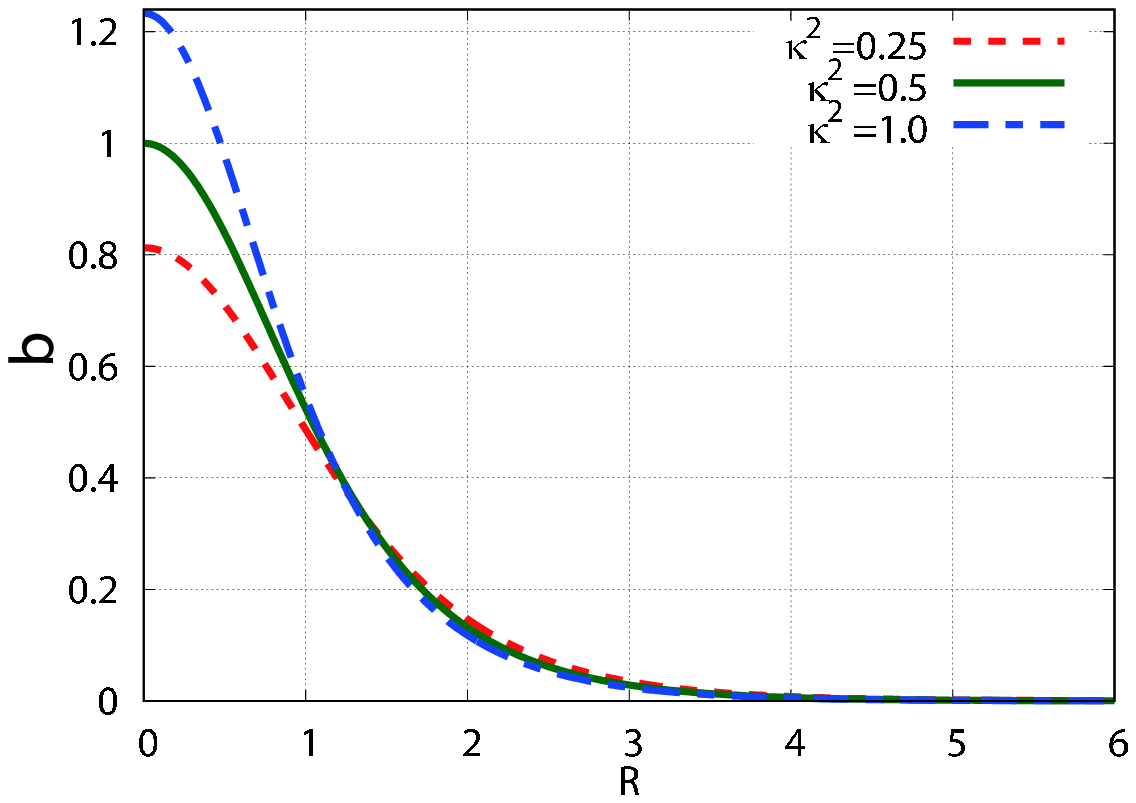}
\caption{(Left panel) The solutions $f$ and $a$ of the field equations (\ref{ANO_eq_f}) and (\ref{ANO_eq_a}) as functions of $R$ for various values of the GL parameter defined in (\ref{GL}), $\kappa := \frac{\lambda}{\sqrt{2} q} = \frac{1}{2} , \frac{1}{\sqrt{2}}$, and $  1$, with a unit winding number $n=1$.
For convenience, we plot $1 - a$ in place of $a$.
(Right panel) The corresponding dimensionless magnetic field $b$ defined in (\ref{ANO_B}) as a function of $R$ for various values of $\kappa$, $\kappa = \frac{1}{2} , \frac{1}{\sqrt{2}},$ and $1$, with a unit winding number $n=1$.
}
\label{ANO_sol}
\end{figure*}

Therefore, we require the solution to satisfy the boundary conditions for $\rho \to \infty$:
\begin{equation}
f (\rho) \xrightarrow{\rho \to \infty} 1 , \ \ \ 
A (\rho) \xrightarrow{\rho \to \infty} \frac{n}{q \rho}
,
\end{equation}
so that the static energy $E$ does not diverge in the long-distance region $\rho \gg 1$.
Indeed, these boundary conditions describe that in the long-distance region, the scalar field $\phi (x)$ goes to its vacuum value $|\phi (\infty)| = v$ and the gauge field $A_{\mu} (x)$ becomes the pure gauge configuration.

In the limit $\rho \to 0$, we assume
\begin{equation}
f (\rho) \xrightarrow{\rho \to 0} 0  , \ \ \ A (\rho) \xrightarrow{\rho \to 0} 0
,
\label{bc0}
\end{equation}
so that the energy $E$ does not have a short-distance divergence.

Now we can clarify the meaning of the integer $n$ by using the boundary conditions.
Let us consider the magnetic flux $\Phi$ passing through the surface $S$ bounded by a circle $C$ with the center at the origin and the large radius $\rho \to \infty$,
\begin{align}
\Phi := & \int_{S} d \sigma^{\mu \nu} \ F_{\mu \nu} 
= \oint_{C = \partial S} d x^{\mu} \ A_{\mu} \nonumber\\
= & \lim_{\rho \to \infty} \int_{0}^{2 \pi} d \varphi \ \rho A (\rho) 
= \lim_{\rho \to \infty} 2 \pi \rho \frac{n}{q \rho}  
= \frac{2 \pi}{q} n
,
\end{align}
which implies that the integer $n$ corresponds to the quantization of the magnetic flux.
By this reason, we call the integer $n$ the {topological charge}, especially the {winding number} of a vortex.

Motivated by the vacuum solution (\ref{vacuum_sol}), we modify the ansatz for the gauge field $A (\rho)$ as
\begin{equation}
A (\rho) = \frac{n}{q \rho} a (\rho)
.
\label{ANO_ansatz_2}
\end{equation}
Moreover, in order to make the field equations dimensionless, we introduce the dimensionless variable:
\begin{equation}
R := q v \rho 
,
\end{equation}
and redefine the profile functions as $f (\rho) = f (R)$ and $a (\rho) = a (R)$.
Thus, the field equations (\ref{pre_eq_f}), (\ref{pre_eq_A}), and (\ref{pre_eq_j}) are rewritten into
\begin{align}
& f^{\prime \prime} (R) + \frac{1}{R} f^{\prime} (R) - \frac{n^{2}}{R^{2}} \bigl[ 1 - a (R) \bigr]^{2} f (R) \nonumber\\
&\hspace{2cm} + \frac{\lambda^{2}}{q^{2}} \bigl[ 1 - f^{2} (R) \bigr] f (R) = 0 , 
\label{ANO_eq_f}\\
& a^{\prime \prime} (R) - \frac{1}{R} a^{\prime} (R) + 2 \bigl[ 1 - a (R) \bigr] f^{2} (R) = 0
,
\label{ANO_eq_a}
\end{align}
where the prime $({}^{\prime})$ stands for the derivative with respect to $R$.
The boundary conditions are also modified as
\begin{align}
f (R) \xrightarrow{R \to 0} & 0 , \ \ \ 
a (R) \xrightarrow{R \to 0}  0 , 
\label{bc_1} \\
f (R) \xrightarrow{R \to \infty} & 1 , \ \ \ 
a (R) \xrightarrow{R \to \infty} 1
\label{bc_2}
.
\end{align}

We have simultaneously solved the field equations (\ref{ANO_eq_f}) and (\ref{ANO_eq_a}) in a numerical way under the boundary conditions (\ref{bc_1}) and (\ref{bc_2}).
The left panel of FIG.\ref{ANO_sol} shows the solutions $f$ and $a$ of the field equations (\ref{ANO_eq_f}) and (\ref{ANO_eq_a}) as functions of $R$ with a unit winding number $n=1$ for various values $\kappa$, $\kappa  = \frac{1}{2} , \frac{1}{\sqrt{2}}$, and $  1$ of the {\it Ginzburg--Landau (GL) parameter}, which is defined by
\begin{equation}
\kappa := \frac{1}{\sqrt{2}} \frac{\lambda}{q}
.
\label{GL}
\end{equation}
For the physical meaning of the GL parameter, see the next section.
This solution is called the {\it Abrikosov--Nielsen--Olesen (ANO) vortex} \cite{NO1,NO2}.
When we introduce the dimensionless magnetic field $b (R)$ and electric current $j (R)$, (\ref{pre_eq_j}) and (\ref{B_ANO}) are rewritten as follows:
\begin{align}
j_{\varphi} (x) := & q^{2} v^{3} j (R) , \ \ \ 
j (R) = \frac{2 n}{R} \bigl[ 1-  a (R) \bigr] f^{2} (R), 
\label{ANO_J} \\
B_{z} (x) := & q v^{2} b (R) , \ \ \ b (R) := \frac{n}{R} a^{\prime} (R)
\label{ANO_B}
.
\end{align}
The right panel of FIG.\ref{ANO_sol} shows the dimensionless magnetic field $b (R)$ corresponding to (\ref{ANO_B}).
Notice that the magnetic field $b (R)$ has no short-distance divergences, which is supported by the boundary condition (\ref{bc0}).
This means that the boundary condition (\ref{bc0}) implies the regularity of the magnetic field $b (R)$ and the finiteness of the energy $E$ for a short distance.

\subsection{Type of the superconductor}

\begin{figure*}[t]
\centering
\includegraphics[width=0.4\textwidth]{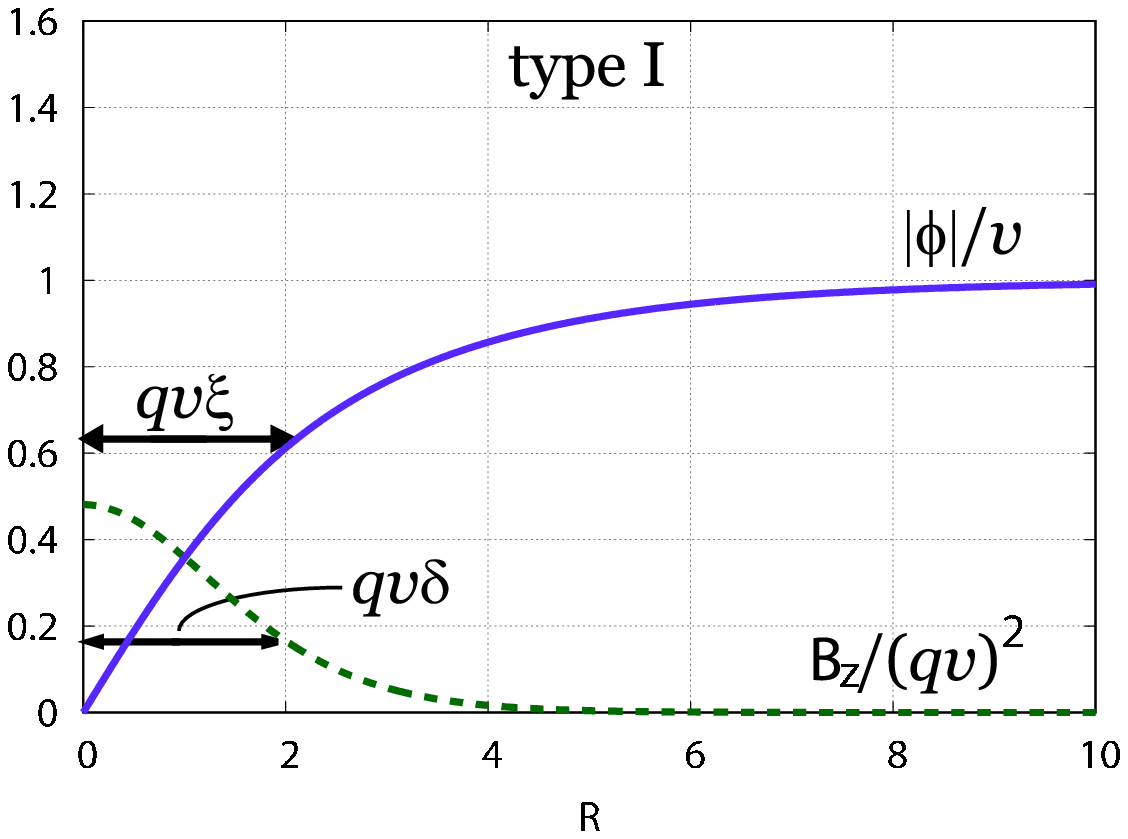} \ \ \  \ \ \ 
\includegraphics[width=0.4\textwidth]{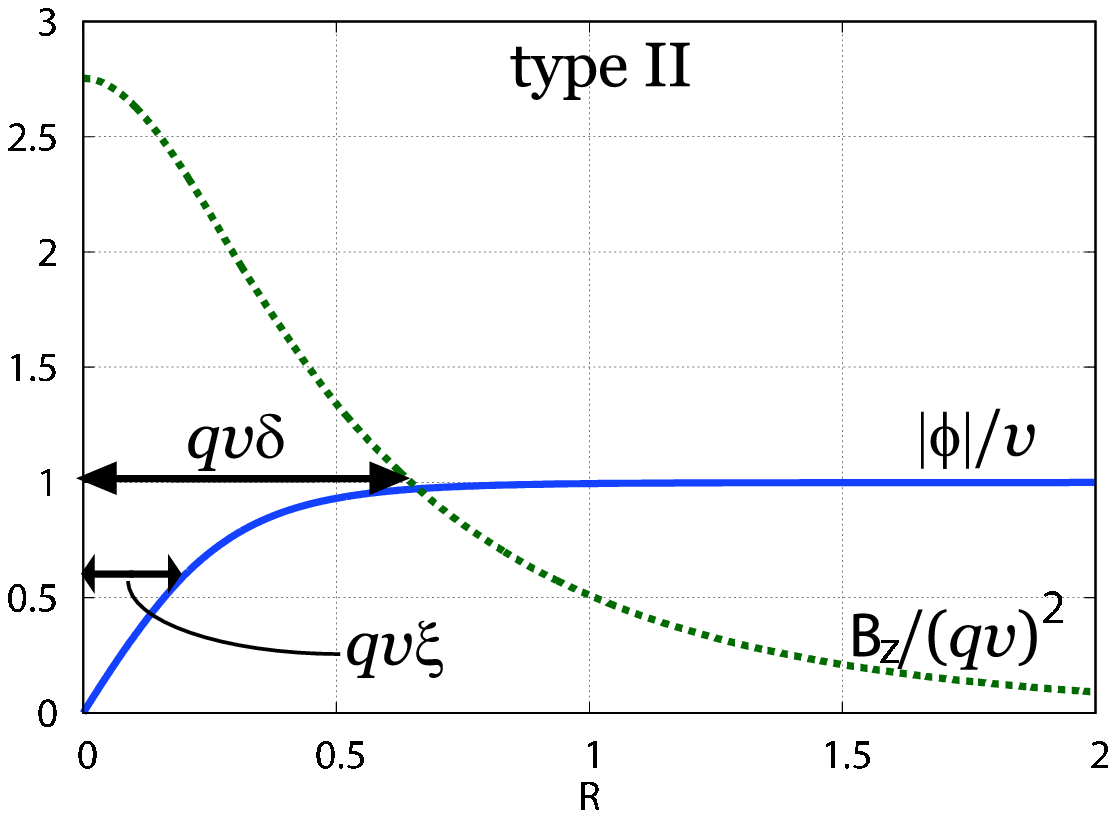}
\caption{The penetration and coherent lengths:
(Left panel) The type I superconductor with $\kappa = \frac{1}{5}$.
(Right panel) The type I\hspace{-.1em}I superconductor with $\kappa = 4$.}
\label{length}
\end{figure*}

In order to investigate the asymptotic forms of the profile functions in the long-distance region $R \gg 1$, we introduce $g$ and $w$ in place of $f$ and $a$ as functions of $R$ by
\begin{equation}
f (R) = 1 - g (R) , \ \ \ a (R) = 1 - R w (R) , 
\end{equation}
where $ |g (R)| , |w (R)| \ll 1$ for $R \gg 1$.
Then, the field equations for $g$ and $w$ read
\begin{align}
& g^{\prime \prime} (R) + \frac{1}{R} g^{\prime} (R) -2 \frac{\lambda^{2}}{q^{2}} g (R) = n^{2} w^{2} (R) , 
\label{eq_g}\\
& w^{\prime \prime} (R) + \frac{1}{R} w^{\prime} (R) - \left( \frac{1}{R^{2}} + 2 \right) w (R) = 0
\label{eq_w}
.
\end{align}
The second equation (\ref{eq_w}) can be solved by using the modified Bessel function of the second kind $K_{\nu} (x)$ as
\begin{equation}
w (R) = C_{1} K_{1} \left( \sqrt{2} R \right) = C_{1} K_{1} \left( \sqrt{2} q v \rho \right)
,
\end{equation}
which behaves for $R \gg 1$ as
\begin{equation}
w (R) \approx C_{1} \sqrt{\frac{\pi}{2 \sqrt{2} R}} e^{- \sqrt{2} R} = C_{1} \sqrt{\frac{\pi}{2 \sqrt{2} q v \rho}} e^{- \sqrt{2} q v \rho}
.
\label{w}
\end{equation}
Therefore, the magnetic field $B_{z} (R)$ has the asymptotic form for $R \gg 1$:
\begin{align}
B_{z} (R) = & q v^{2} \frac{n}{R} \frac{d}{d R} \bigl[ 1 - R w (R) \bigr] \nonumber\\
= &  q v^{2} C_{1} n \sqrt{2} K_{0} \left( \sqrt{2} R \right) \nonumber\\
\approx & q v^{2} C_{1} n \sqrt{2} \sqrt{\frac{\pi}{2 \sqrt{2} R}} e^{- \sqrt{2} R} \nonumber\\
= & q v^{2} C_{1} n \sqrt{\frac{\pi}{\sqrt{2} q v \rho}} e^{- \sqrt{2} q v \rho}
,
\end{align}
where we have used the formula $z K_{\nu}^{\prime} (z) + \nu K_{\nu} (z) = - z K_{\nu - 1} (z)$.

Inserting the asymptotic form (\ref{w}) of $w (R)$ into the first equation (\ref{eq_g}), we have the closed equation for $g (R)$
\begin{equation}
g^{\prime \prime} (R) + \frac{1}{R} g^{\prime} (R) - 2 \frac{\lambda^{2}}{q^{2}} g (R) = n^{2} C_{1}^{2} \frac{\pi}{2 \sqrt{2} R} e^{- \sqrt{2} R}
.
\label{eq_asymp_g}
\end{equation}
The solution of this inhomogeneous equation is given by
\begin{equation}
g (R) = C_{2} K_{0} \left( \sqrt{2} \frac{\lambda}{q} R \right) + \frac{\pi n^{2} C_{1}^{2}}{2 \sqrt{2} \left( 8 - \frac{\lambda^{2}}{q^{2}} \right)} \frac{1}{R} e^{- 2 \sqrt{2} R}
,
\end{equation}
where the first term is the general solution of the homogeneous equation obtained by ignoring the right hand side of (\ref{eq_asymp_g}) and the second term is a particular solution of (\ref{eq_asymp_g}).
In terms of the dimensionful variable $\rho$, $g (R)$ behaves as
\begin{align}
g (R) = \left\{ \begin{array}{cc}
C_{2} \sqrt{\frac{\pi}{2 \sqrt{2} \lambda v \rho}} e^{- \sqrt{2} \lambda v \rho} & \left( \frac{\lambda}{q} \leq 2 \sqrt{2} \right) \\
\frac{\pi n^{2} C_{1}^{2}}{2 \sqrt{2} q v \left( 8 - \frac{\lambda^{2}}{q^{2}} \right)} \frac{1}{\rho} e^{- 2 \sqrt{2} q v \rho} & \left( \frac{\lambda}{q} > 2 \sqrt{2} \right)
\end{array} \right.
,
\end{align}
which means that the fall-off factor of the scalar field must be distinguished by the value of $\lambda / q$.

We can define two typical lengths $\delta$ and $\xi$ by
\begin{equation}
\delta := \frac{1}{\sqrt{2} q v} = \frac{1}{m_{V}} , \ \ \ \xi := \frac{1}{\lambda v} = \frac{\sqrt{2}}{m_{S}}
,
\label{penetration_coherent}
\end{equation}
and the ratio by
\begin{equation}
\kappa := \frac{\delta}{\xi} = \frac{1}{\sqrt{2}} \frac{m_{S}}{m_{V}} = \frac{1}{\sqrt{2}} \frac{\lambda}{q}
.
\end{equation}
The length $\delta$ is called the {\it penetration length} (or depth), at which the magnitude of the magnetic field $B_{z}$ falls to $1/e \simeq 37 \%$ of its original value at the origin $\rho = 0$.
The length $\xi$ is called the {\it coherent length} because the magnitude of the scalar field $|\phi (x)|$ grows to $1 - 1/e \simeq 63 \%$ of its vacuum value $v$. 
(See FIG.\ref{length}.)
Taking into account the fall-off rates (or the masses) of the gauge and scalar fields, the mass of the gauge field $m_{V} = \sqrt{2} q v$ is larger than that of the scalar field $m_{S} = \sqrt{2} \lambda v$ for $\kappa < \frac{1}{\sqrt{2}}$, while for $\kappa > \frac{1}{\sqrt{2}}$ the opposite situation occurs.
At the critical value $\kappa = \frac{1}{\sqrt{2}}$, the two masses $m_{V}$ and $m_{S}$ become equal: $m_{V} = m_{S}$.
Therefore, the superconductor is classified by the value of the ratio $\kappa$ as
\begin{equation}
\kappa < \frac{1}{\sqrt{2}}: \ {\rm type \ I} , \ \ \ 
\kappa = \frac{1}{\sqrt{2}}: \ {\rm BPS} , \ \ \ 
\kappa > \frac{1}{\sqrt{2}}: \ {\rm type \ I\hspace{-.1em}I}
.
\end{equation}
The ratio $\kappa$ is called the {\it Ginzburg--Landau (GL) parameter}.
The limit $\kappa \to \infty$, which is realized by $\xi \to 0$ or $m_{S} \to \infty$, is called the {\it London limit}.

\section{Type of dual superconductor}

To determine the type of dual superconductivity for $SU(2)$ Yang--Mills theory, we simultaneously fit the chromoelectric field and the induced magnetic current obtained by the lattice simulation \cite{Kato-Kondo-Shibata} (see FIG.\ref{lattice_result_V} and FIG.\ref{lattice_current}) to the magnetic field and electric current of the $n=1$ ANO vortex.

\subsection{The previous study using the Clem ansatz}

In this subsection, we give a review of the approximated method of fitting with the Clem ansatz \cite{Clem}.
The previous studies \cite{Cea-Cosmai1,Cea-Cosmai2, type1-1,type1-2, Kato-Kondo-Shibata, Shibata-Kondo-Kato-Shinohara} considered only the regression of the chromoelectric flux, however in this paper, we also take into account the regression of the induced magnetic current to compare with our new method. 
In the Clem ansatz adopted to the $U(1)$ gauge-scalar model, the scalar profile function $f (\rho)$ is assumed to be
\begin{equation}
f (\rho) = \frac{\rho}{\sqrt{\rho^{2} + \zeta^{2}}}
,
\label{Clem_assump}
\end{equation}
where $\zeta$ is a variational parameter for the core radius of the ANO vortex and $\rho$ is a dimensionful variable $\rho = R/(q v)$.
For the profile function of the gauge field $a (\rho)$, we introduce a new function $w (\rho)$ by
\begin{equation}
a (\rho) = 1 - \frac{\sqrt{\rho^{2} + \zeta^{2}}}{\zeta} \frac{ w (\sqrt{\rho^{2} + \zeta^{2}})}{ w (\zeta)}
,
\end{equation}
which satisfies the boundary condition (\ref{bc_1}), i.e., $a (\rho = 0) = 0$.
Then, the field equation (\ref{ANO_eq_a}) for the gauge field is now written as the differential equation for $w$:
\begin{equation}
\frac{d^{2} w (x)}{d x^{2}} + \frac{1}{x} \frac{d w (x)}{d x} - \left( \frac{1}{x^{2}} + 2 q^{2} v^{2} \right) w (x) = 0
,
\end{equation}
where we have defined a variable $x := \sqrt{\rho^{2} + \zeta^{2}}$.
The solution is given by the modified Bessel function of the second kind $K_{\nu} (z)$ as
\begin{equation}
w (x) \propto K_{1} \left( \sqrt{2} q v x \right)
,
\end{equation}
and hence
\begin{equation}
a (\rho) = 1 - \frac{\sqrt{\rho^{2} + \zeta^{2}}}{\zeta} \frac{K_{1} \left(\sqrt{2} q v \sqrt{\rho^{2} + \zeta^{2}}\right)}{K_{1} (\sqrt{2} q v \zeta)}
.
\end{equation}
Therefore, the magnetic field $B (\rho)$ is given by
\begin{equation}
B (\rho) =  \alpha K_{0} \left( \beta \sqrt{\rho^{2} + \zeta^{2}}\right)
,
\end{equation}
where we have defined
\begin{equation}
\beta := \sqrt{2} q v , \ \ \ \alpha := \frac{\Phi}{2\pi} \frac{\beta}{\zeta} \frac{1}{K_{1} \left( \beta \zeta\right)}
,
\end{equation}
with the external flux $\Phi = 2 \pi n /q$.
The electric current $\bm{J} (\rho) = J (\rho) \bm{e}_{\varphi}$ is also written as:
\begin{align}
J (\rho)
= & \alpha \beta \frac{\rho}{\sqrt{\rho^{2} + \zeta^{2}}} K_{1} \left( \beta \sqrt{\rho^{2} + \zeta^{2}} \right)
.
\end{align}
In the present setting, the energy per unit length $E$ can be calculated by restricting ourselves to the unit vortex with $n = 1$ as
\begin{align}
E = & 2 \pi v^{2} \biggl[ \frac{1}{4} + \frac{1}{4} s^{2} \kappa^{2} + \frac{1}{s} \frac{K_{0} (s)}{K_{1} (s)} \biggr]
,
\label{Clem_energy}
\end{align}
where we have introduced the parameter $s = \sqrt{2} q v \zeta$.
Since the vortex solution is obtained by minimizing the energy with respect to the parameter $s$, or $\zeta$, for a given GL parameter $\kappa$, the energy (\ref{Clem_energy}) must satisfy
\begin{equation}
0 = \frac{d}{d s} \frac{E}{2 \pi v^{2}} = \frac{1}{2} \kappa^{2} s - \frac{1}{s} + \frac{1}{s} \left( \frac{K_{0} (s)}{K_{1} (s)} \right)^{2} 
.
\end{equation}
Therefore, the GL parameter $\kappa$ is given by
\begin{align}
\kappa = & \frac{\sqrt{2}}{s} \sqrt{1 - \left( \frac{K_{0} (s)}{K_{1} (s)} \right)^{2}} \nonumber\\
= & \frac{\sqrt{2}}{\sqrt{2} q v \zeta} \sqrt{1 - \left( \frac{K_{0} (\sqrt{2} q v \zeta)}{K_{1} (\sqrt{2} q v \zeta)} \right)^{2}}
.
\label{GL_Clem}
\end{align}

\begin{figure*}[t]
\centering
\includegraphics[width=0.45\textwidth]{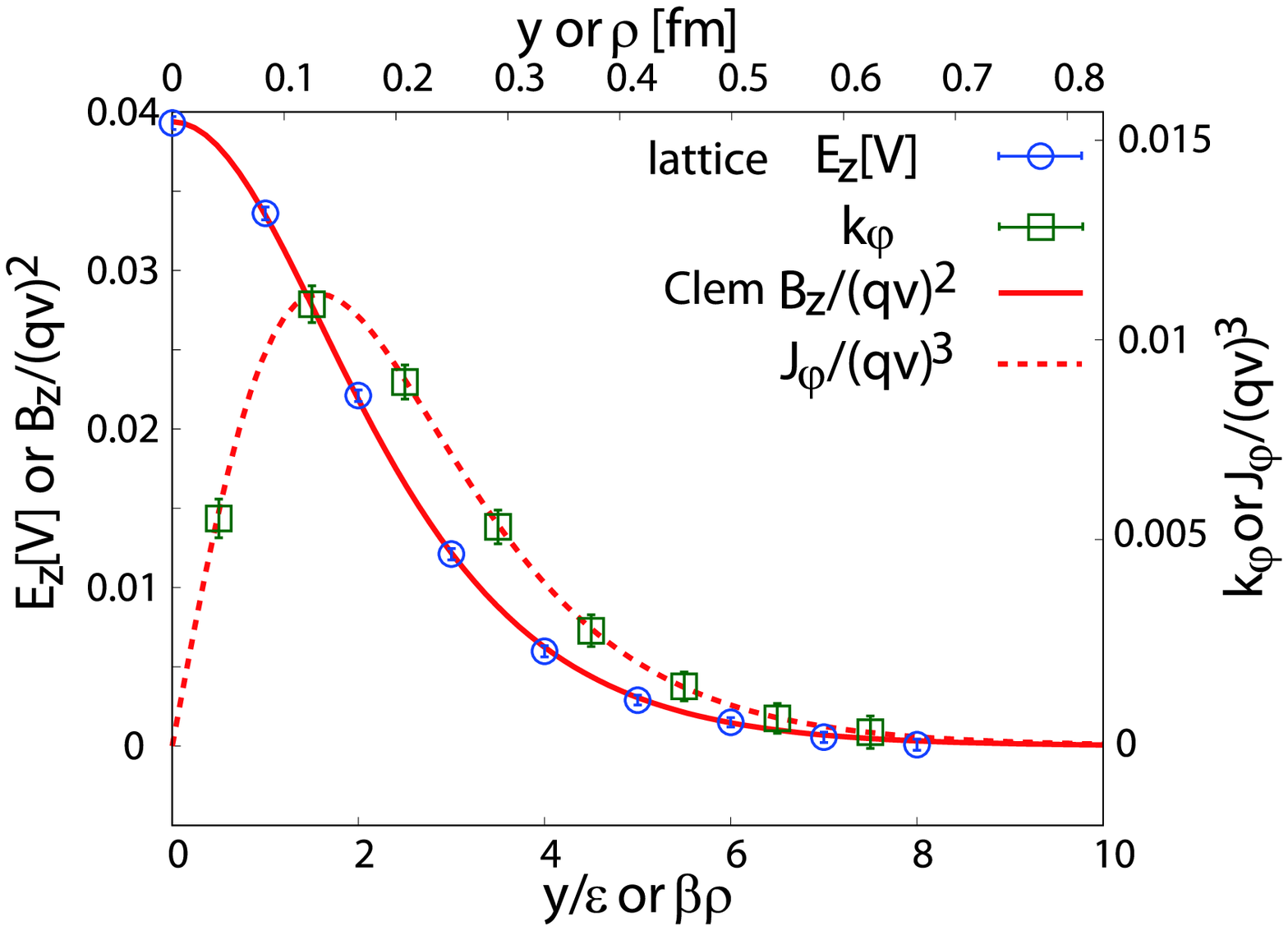} \ \ \ 
\includegraphics[width=0.45\textwidth]{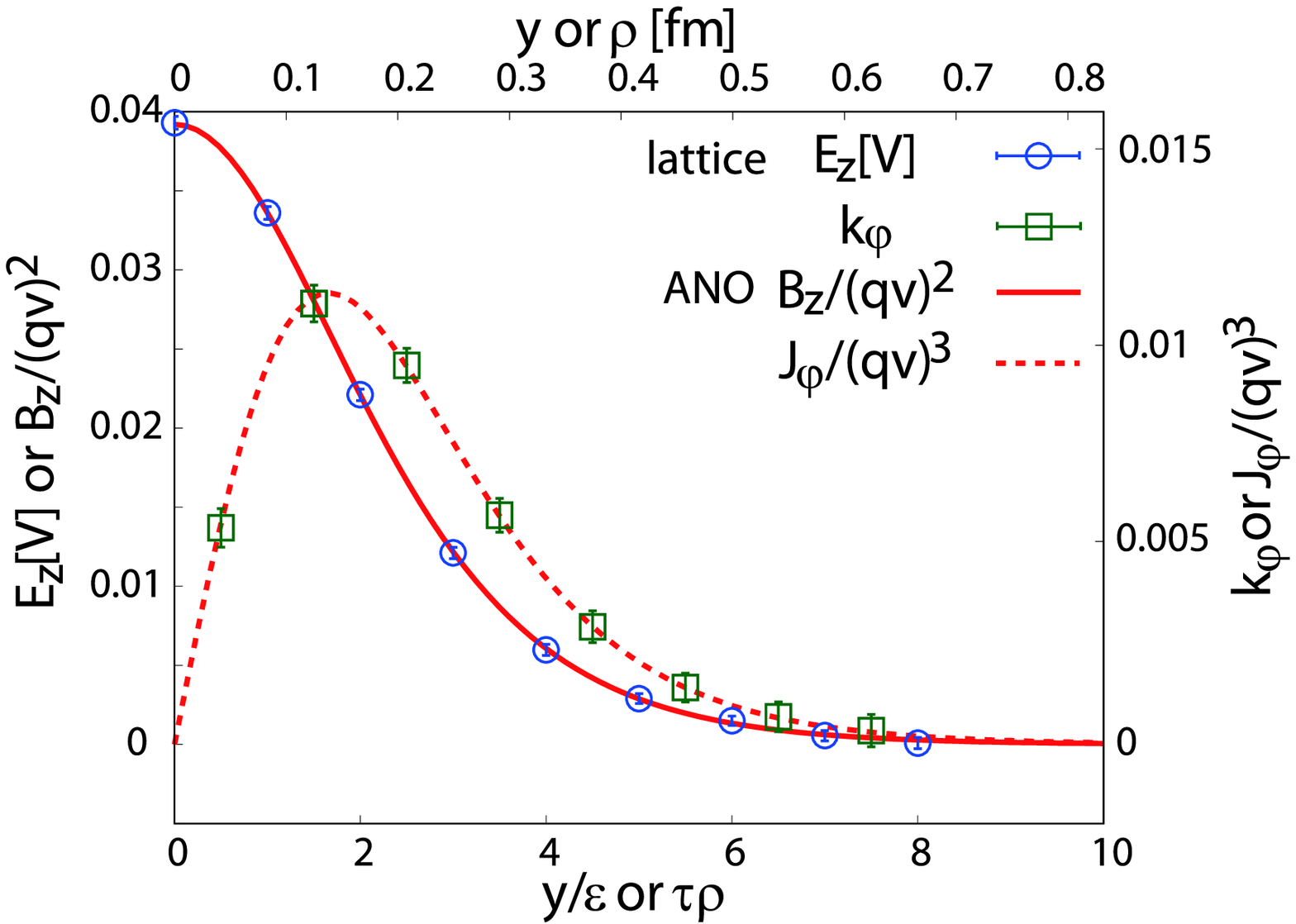} 
\caption{The fitting results: (right panel) the approximated method based on the Clem ansatz including both the flux and magnetic current (\ref{Clem_4}), (left panel) the new method (\ref{fit_result}) by solving the field equations of the ANO vortex with a unit winding number.}
\label{fit_fig}
\end{figure*}

In the previous study \cite{Kato-Kondo-Shibata}, we adopted the fitting only for the chromoelectric flux.
In this paper, we adopt the fitting for the chromoelectric flux and magnetic current simultaneously.
In what follows, we use values measured in the lattice unit, e.g., the distance $\hat{y} = y/\epsilon$ with a lattice spacing $\epsilon$, the chromoelectric flux $E_{z} (\hat{y}) = F_{34} [V] (\hat{y})$ in (\ref{strength_V}), and the magnetic current $k_{\varphi} (\hat{y})$ in (\ref{current_lattice}).
Then, we denote the set of data as $(\hat{y}_{i} , E_{z} (\hat{y}_{i}) , \delta E_{z} (\hat{y}_{i}))$ for the chromoelectric field, and $(\hat{y}_{j} , k_{\varphi} (\hat{y}_{j}) , \delta k_{\varphi} (\hat{y}_{j}))$ for the induced magnetic current, where $\delta \mathcal{O}$ represents the error of the measurement $\mathcal{O}$.

To define dimensionless regression functions, let us rescale the parameters $\beta$ and $\zeta$ to be dimensionless by using the lattice spacing $\epsilon$ as
\begin{equation}
\hat{\beta} := \beta \epsilon , \ \ \ 
\hat{\zeta} := \frac{\zeta}{\epsilon} 
,
\end{equation}
and hence a parameter $\alpha$ is rescaled as
\begin{equation}
\hat{\alpha} := \alpha \epsilon^{2}
.
\end{equation}
We also rescale the magnetic field $B$ and the electric current $J$ as
\begin{equation}
\hat{B} := \epsilon^{2} B , \ \ \ 
\hat{J} := \epsilon^{3} J
.
\end{equation}
Then, we can define regression functions by
\begin{align}
\hat{B} (\hat{\rho}; \hat{\alpha}, \hat{\beta} , \hat{\zeta}) = & \hat{\alpha} K_{0} \left( \hat{\beta} \sqrt{\hat{\rho}^{2} + \hat{\zeta}^{2}} \right)
, \\
J (\hat{\rho}; \hat{\alpha}, \hat{\beta}, \hat{\zeta}) = & \hat{\alpha} \hat{\beta} \frac{\hat{\rho}}{\sqrt{\hat{\rho}^{2} + \hat{\zeta}^{2}}} K_{1} \left( \hat{\beta} \sqrt{\hat{\rho}^{2} + \hat{\zeta}^{2}} \right)
,
\end{align}
with the dimensionless variable $\hat{\rho} := \rho/\epsilon$ in the lattice unit.
Then, the error functions of the regression with the weights are given by
\begin{align}
\varepsilon_{\rm flux} ( \hat{y}_{i} ;  \hat{\alpha} , \hat{\beta} , \hat{\zeta} ) = & \frac{E_{z} (\hat{y}_{i}) - \hat{B} (\hat{y}_{i}; \hat{\alpha} , \hat{\beta} , \hat{\zeta})}{\delta E_{z} (\hat{y}_{i}) } , \label{Clem_error_flux}\\
\varepsilon_{\rm current} (\hat{y}_{j} ; \hat{\alpha} , \hat{\beta} , \hat{\zeta}) = & \frac{k_{\varphi} (\hat{y}_{j}) - \hat{J} (\hat{y}_{j} ; \hat{\alpha},  \hat{\beta} , \hat{\zeta})}{\delta k_{\varphi} (\hat{y}_{j})}
.\label{Clem_error_current}
\end{align}
When we assume that these errors  obey independent standard normal distributions, the parameters $\hat{\alpha} , \hat{\beta}$, and $\hat{\zeta}$ can be estimated by maximizing the log-likelihood function $\ell (\hat{\alpha}, \hat{\beta}, \hat{\zeta})$ for (\ref{Clem_error_flux}) and (\ref{Clem_error_current}) defined by
\begin{align}
 \ell ( \hat{\alpha} , \hat{\beta} , \hat{\zeta} )  = &- \frac{1}{2} \sum_{i = 1}^{n} \left( \varepsilon_{\rm flux} (\hat{y}_{i} ; \hat{\alpha} , \hat{\beta} , \hat{\zeta} ) \right)^{2} \nonumber\\
& - \frac{1}{2} \sum_{j = 1}^{m} \left( \varepsilon_{\rm current} (\hat{y}_{j} ; \hat{\alpha},  \hat{\beta} , \hat{\zeta} ) \right)^{2}
.
\label{fitting}
\end{align}

The GL parameter $\kappa$ is determined according to (\ref{GL_Clem}) in terms of the estimated values $\hat{\beta}_{\star}$ and $\hat{\zeta}_{\star}$ by
\begin{equation}
\kappa_{\star} = \frac{\sqrt{2}}{\hat{\beta}_{\star} \hat{\zeta}_{\star}} \sqrt{1 - \left( \frac{K_{0} (\hat{\beta}_{\star} \hat{\zeta}_{\star})}{K_{1} (\hat{\beta}_{\star} \hat{\zeta}_{\star})} \right)^{2}}
.
\end{equation}

The obtained values in the previous work \cite{Kato-Kondo-Shibata}, which can be achieved by ignoring the second term in (\ref{fitting}) and restricting the fitting range to $2 \leq \hat{\rho} \leq 8$, are given by
\begin{align}
&\hat{\alpha}_{\star} = 0.41 \pm 0.44 , \ \ \ 
\hat{\beta}_{\star} = 0.77 \pm 0.13  , \nonumber\\
&\hat{\zeta}_{\star} = 2.75 \pm 0.79 , \ \ \ 
\kappa_{\star} = 0.38 \pm 0.23 , \label{Clem1}\\
&{\rm MSR}_{\rm flux} := \sum_{i} \epsilon^{2}_{\rm flux} (\hat{y}_{i}; \hat{\alpha}_{\star} , \hat{\beta}_{\star}, \hat{\zeta}_{\star} ) /{\rm d.o.f.} = 0.171 \nonumber
,
\end{align}
where ${\rm MSR}_{\rm flux}$ is the sum of squared residuals for the regression of (\ref{Clem_error_flux}) divided by the degrees of freedom (d.o.f.) for fitting: (the number of  data points) minus (the number of independent variational parameters), i.e., ${\rm d.o.f.} = 7 - 3 = 4$.

By incorporating also the regression of the electric current $J$, the fitting result is in good agreement with (\ref{Clem1}):
\begin{align}
&\hat{\alpha}_{\star} = 0.43 \pm 0.42 , \ \ \ 
\hat{\beta}_{\star} = 0.78 \pm 0.12, \nonumber\\
& \hat{\zeta}_{\star} = 2.78 \pm 0.70 , \ \ \ 
\kappa_{\star} = 0.37 \pm 0.20 ,\label{Clem_2}\\
&{\rm MSR}_{\rm flux} = 0.171 , \ \ \ 
{\rm MSR}_{\rm current} = 0.086 , \nonumber\\
&{\rm MSR}_{\rm total} = 0.135 \nonumber
.
\end{align}
It should be noticed that the fitting range is restricted to $2 \leq \hat{\rho} \leq 8$ as well as (\ref{Clem1}).

We further investigate the fitting by using the whole range $0 \leq \hat{\rho} \leq 8$.
The result obtained by using only the flux is given by
\begin{align}
&\hat{\alpha}_{\star} = 0.58 \pm 0.31 , \ \ \ 
\hat{\beta}_{\star} = 0.811 \pm 0.070  , \nonumber\\
&\hat{\zeta}_{\star} = 3.00 \pm 0.30 , \ \ \ 
\kappa_{\star} = 0.315 \pm 0.080 , \label{Clem3}\\
&{\rm MSR}_{\rm flux} = 0.133 \nonumber
.
\end{align}
By including the magnetic current, the result is given by
\begin{align}
&\hat{\alpha}_{\star} = 0.63 \pm 0.32 , \ \ \ 
\hat{\beta}_{\star} = 0.821 \pm 0.066, \nonumber\\
& \hat{\zeta}_{\star} = 3.05 \pm 0.28 , \ \ \ 
\kappa_{\star} = 0.303 \pm 0.071 ,\label{Clem_4}\\
&{\rm MSR}_{\rm flux} = 0.141 , \ \ \ 
{\rm MSR}_{\rm current} = 0.106 , \nonumber\\
&{\rm MSR}_{\rm total} = 0.125 \nonumber
.
\end{align}
The fitting result (\ref{Clem_4}) is shown in the left panel of FIG.\ref{fit_fig}.
We find that the inclusion of the short range modifies the value of the GL parameter $\kappa$ to a smaller one.
We also find that the inclusion of the regression for the magnetic current indeed improves the accuracy of the fitting in both cases of the fitting range $2 \leq \hat{\rho} \leq 8$ and $0 \leq \hat{\rho} \leq 8$.

\subsection{The new method}

In this subsection, we shall fit the chromoelectric flux and the magnetic current to the magnetic field and the electric current of the ANO vortex simultaneously {\it without any approximations}.
The advantage of the new method could be that the value of the GL parameter $\kappa$ is a direct fitting parameter unlike the case in the Clem ansatz. 

Such a fitting can be done by using the regression functions $B$ and $J$ constructed by the solutions, $f (R)$ and $a (R)$, of the field equations (\ref{ANO_eq_f}) and (\ref{ANO_eq_a}) through the dimensionless magnetic field $b (R)$ in (\ref{ANO_B}) and the electric current $j (R)$ in (\ref{ANO_J}).
However, there are difficulties to estimate the model parameters, when we flow the same procedure as in the previous subsection.
When we construct the regression functions $B$ and $J$ from the numerical solutions $f (R)$ and $a (R)$ by solving the field equations (\ref{ANO_eq_f}) and (\ref{ANO_eq_a}), we also calculate the regression functions numerically.
Indeed, it is necessary to numerically calculate the derivative in (\ref{ANO_B}) separately, and this causes a large numerical error even if one obtains the solutions $f (R)$ and $a (R)$ with small errors.
To avoid these difficulties, we reorganize the field equations to include both $b (R)$ and $j (R)$ as independent unknown functions by
\begin{align}
& f^{\prime \prime} (R) + \frac{1}{R} f^{\prime} (R) - \frac{n^{2}}{R^{2}} \bigl[ 1- a (R) \bigr]^{2} f (R)  \nonumber\\
&\hspace{2cm}  + 2 \kappa^{2} \bigl[ 1 - f^{2} (R) \bigr] f (R) = 0 , \label{fit_eq_f}\\
& b^{\prime} (R) +  j (R)  = 0 , \label{eq_b} \\
& n a^{\prime} (R) =  R b (R) , \label{eq_a}\\
& j (R) =  \frac{2n}{R} \bigl[ 1 - a (R) \bigr] f^{2} (R)
,\label{eq_j}
\end{align}
where we have decomposed the second order differential equation (\ref{ANO_eq_a}) for the gauge profile function $a (R)$ into two independent first order differential equations (\ref{eq_b}) and (\ref{eq_a}) and one algebraic equation (\ref{eq_j}).
We solve these coupled equations simultaneously.
We impose the following boundary conditions for four unknown functions $f (R) , a (R) , b (R)$, and $j (R)$:
\begin{align}
f (0) = & 0 , \ \ \ b^{\prime} (0) =  0 ,\\
f ( \infty) =  &1  , \ \ \  a (\infty) = 1
.
\end{align}

From (\ref{ANO_J}) and (\ref{ANO_B}) , we obtain the regression functions with the dimensionless variational parameters $\hat{\eta} := q v^{2} \epsilon^{2}$  $\hat{\tau} := q^{2} v^{3} \epsilon^{3}$ in the lattice unit by
\begin{align}
\hat{B} (\hat{\rho} ; \hat{\eta} ,\hat{\tau} , \kappa) :=  \hat{\eta} b (\hat{\tau} \hat{\rho}; \kappa) , \ \ \ 
\hat{J} (\hat{\rho} ; \hat{\eta} ,\hat{\tau} , \kappa) :=  \hat{\eta} \hat{\tau} j (\hat{\tau} \hat{\rho}; \kappa)
,
\label{new_res}
\end{align}
where $\hat{\rho} := \rho/\epsilon$ is a dimensionless variable, and $\kappa$ is the GL parameter.

By numerically solving (\ref{fit_eq_f})--(\ref{eq_j}) simultaneously and maximizing the log-likelihood function (\ref{fitting}) with the regression functions (\ref{new_res}) by varying the parameters $\hat{\eta} , \hat{\tau}$, and $\kappa$, we estimate the model parameters $\hat{\eta}, \hat{\tau}$, and $\kappa$.
Note that since the coupled differential equations (\ref{fit_eq_f})--(\ref{eq_j}) with respect to $R = \hat{\tau} \hat{\rho}$ depends on only the GL parameter $\kappa$, the variation of the parameters $\hat{\tau}$ and $\hat{\eta}$ does not deform the functions $b (R)$ and $j (R)$.
Thus we obtain the results:
\begin{align}
& \hat{\eta}_{\star} = 0.0448 \pm 0.0050 , \ \ \ 
\hat{\tau}_{\star} = 0.508 \pm 0.032  , \nonumber\\
& \kappa_{\star} = 0.565 \pm 0.053 , \label{fit_result}\\
&{\rm MSR}_{\rm flux} = 0.131 , \ \ \ 
{\rm MSR}_{\rm current} = 0.0938 , \nonumber\\
&{\rm MSR}_{\rm total} = 0.114 \nonumber
.
\end{align}
The fitting result is shown in the right panel of FIG.\ref{fit_fig}.

\begin{figure}[t]
\centering
\includegraphics[width=0.4\textwidth]{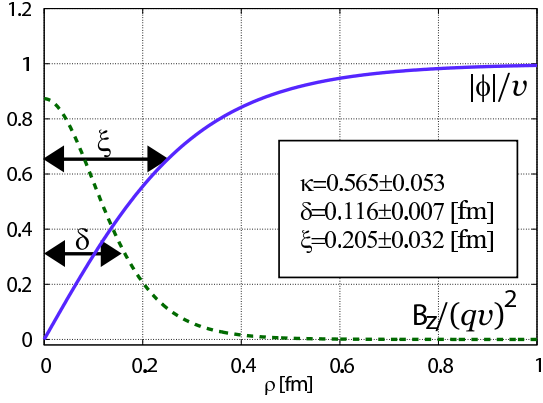}
\caption{The penetration and coherent lengths for the value of the fitted GL parameter $\kappa = 0.565$.}
\label{penet_coh}
\end{figure}

We further obtain the penetration length $\delta$ and coherent length $\xi$ defined in (\ref{penetration_coherent}) by using the fitted values (\ref{fit_result}) as
\begin{align}
\delta = & \frac{\epsilon}{\hat{\tau}_{\star}} = 0.116 \pm 0.007 \ \mathrm{fm} , \\
\xi = & \frac{\delta}{\kappa_{\star}} = 0.205 \pm 0.032 \ \mathrm{fm}
.
\end{align}
FIG.\ref{penet_coh} shows the penetration and coherent lengths for the fitted value of the GL parameter $\kappa$ with corresponding functions $|\phi|/v$ and $B_{z} /(qv)^{2}$.
See also FIG.\ref{length}.

This new result shows that the vacuum of $SU(2)$ Yang--Mills theory is of type I, $\kappa = 0.565 \pm 0.053 < 1/\sqrt{2} \approx 0.707$, which is consistent with the results based on  the Clem ansatz (\ref{Clem1}) and (\ref{Clem_2}) within errors.
We find that the inclusion of the regression for the magnetic current (\ref{Clem_2}), (\ref{Clem_4}), and (\ref{fit_result}) give small errors of the GL parameter $\kappa$ than the excluded ones (\ref{Clem1}) and (\ref{Clem3}).
We also observe that the sums of squared residuals for both the flux and current in the new method become smaller than the fitting method based on the Clem ansatz.
Therefore, the inclusion of the fitting for the magnetic current is important to improve the accuracy.

We should be aware of the effect of changing the fitting range.
If we choose the fitting range $2 \leq \hat{\rho} \leq 8$, the value of the GL parameter $\kappa$ agrees with the result of  $0 \leq \hat{\rho} \leq 8$ within their errorbars.
This fact quite differs from the previous result based on the Clem ansatz represented in (\ref{Clem1})--(\ref{Clem_4}).

\section{Distribution of the stress force around a vortex}

In what follows, to clarify the difference between type I and I\hspace{-.1em}I of dual superconductors in view of force among the chromoelectric fluxes, we investigate the Maxwell stress tensor according to the proposal \cite{EMT2-1,EMT2-2,EMT2-3}.
We find that the components  (\ref{EMT_ANO})  of the energy-momentum-stress tensor $T^{\mu \nu}$ around an ANO vortex are written under the ansatz (\ref{ANO_ansatz}), (\ref{ANO_ansatz_2}) and (\ref{ANO_B}) as
\begin{align}
T^{zz} = & q^{2} v^{4} \biggl[ \frac{1}{2} b^{2} (R) + f^{\prime 2} (R) + \frac{n^{2}}{R^{2}} \left( 1 - a (R) \right)^{2} f^{2} (R) \nonumber\\
&\hspace{1cm} + \kappa^{2} \left( 1 - f^{2} (R) \right)^{2} \biggr] = - T^{00} ,\label{Tzz} \\
T^{\rho\rho} = & q^{2} v^{4} \biggl[\frac{1}{2} b^{2} (R) +f^{\prime 2} (R) - \frac{n^{2}}{R^{2}} \left( 1 - a (R) \right)^{2} f^{2} (R) \nonumber\\
& \hspace{1cm}  - \kappa^{2} \left( 1 - f^{2} (R) \right)^{2} \biggr] , \\
T^{\varphi\varphi} = & q^{2} v^{4} \biggl[ \frac{1}{2} b^{2} (R) - f^{\prime 2} (R) + \frac{n^{2}}{R^{2}} \left( 1 - a (R) \right)^{2} f^{2} (R) \nonumber\\
& \hspace{1cm} - \kappa^{2} \left( 1 - f^{2} (R) \right)^{2} \biggr]
,
\end{align}
and all the off-diagonal components vanish.\footnote{
Here, we change the sign of $T^{j k}$ defined in (\ref{EMT_ANO}) by using the ambiguity of the overall sign of the Noether current in order to reproduce the conventional Maxwell stress tensor.}

\begin{figure}[t]
\centering
\includegraphics[width=0.4\textwidth]{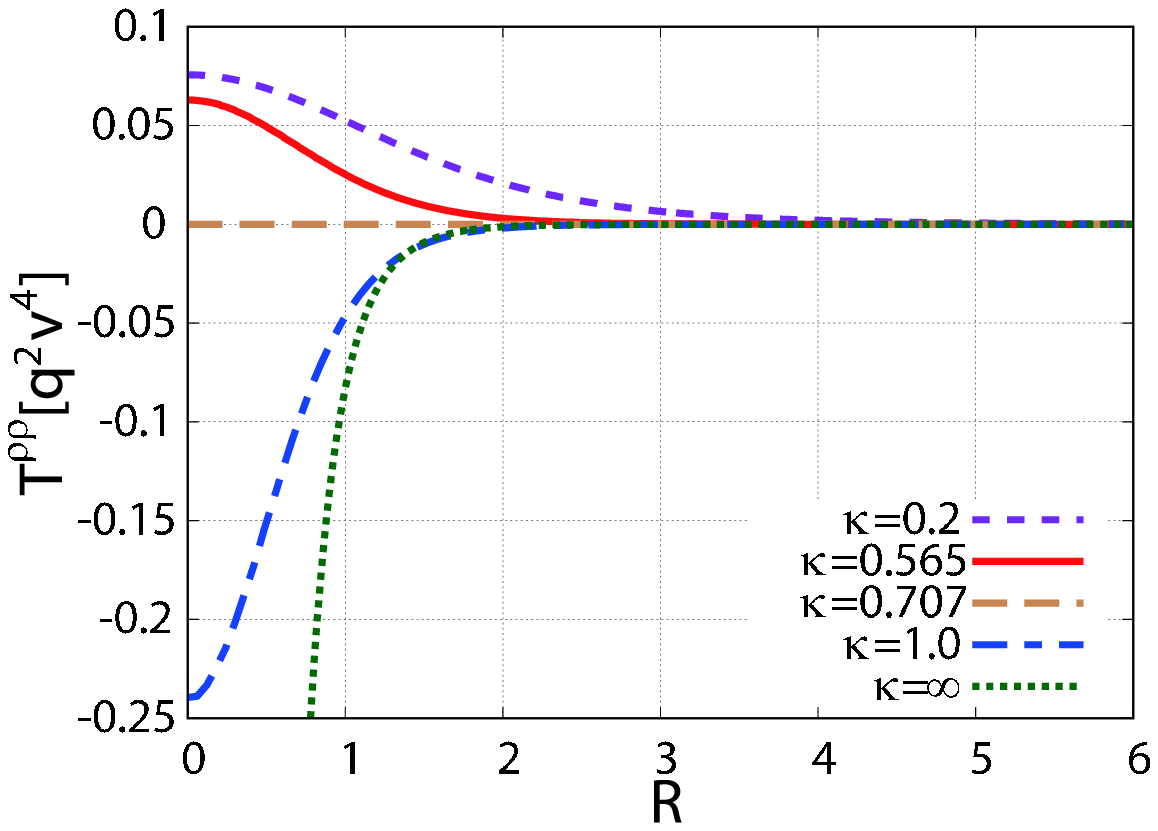} \\ \includegraphics[width=0.4\textwidth]{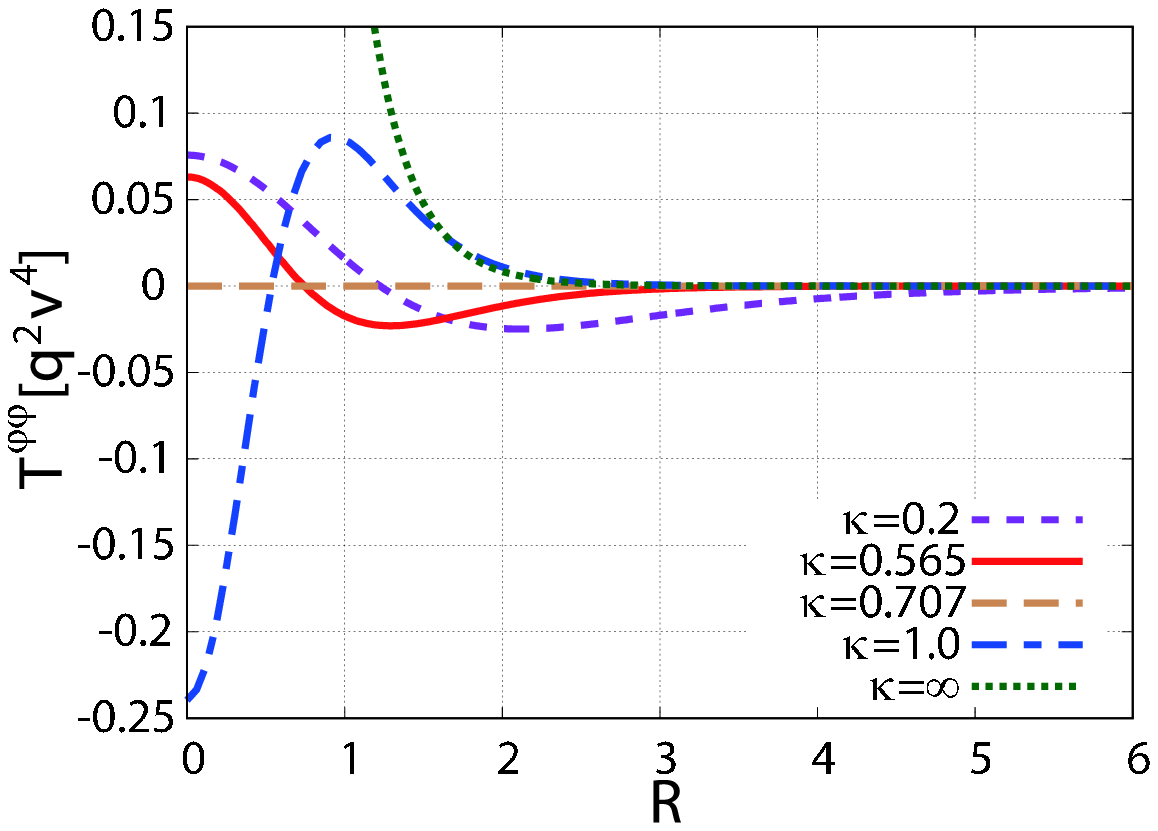} \\  
\includegraphics[width=0.4\textwidth]{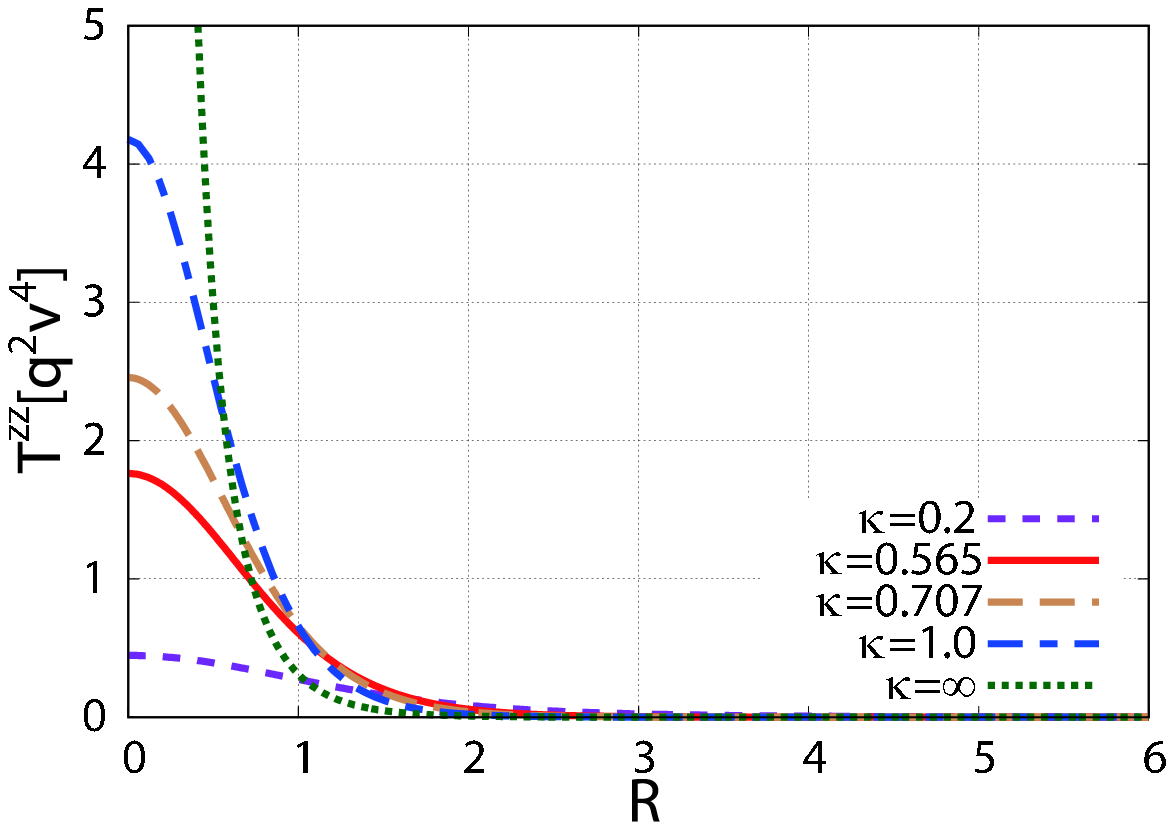} 
\caption{The components of the stress tensor $T^{\mu \nu}$ as functions of $R$ for the $n=1$ ANO vortex configuration in the unit of $q^{2} v^{4}$ for $\kappa = \frac{1}{5} , 0.565$ (type I), $\frac{1}{\sqrt{2}}$ (BPS), $1$ (type I\hspace{-.1em}I), and $\infty$ (London limit).
The red solid curves represent the stress tensor for the fitted parameter of the GL parameter $\kappa = 0.565$:  $T^{\rho\rho}$ (top panel), $T^{\varphi \varphi}$ (middle panel), and $T^{zz}$ (bottom panel)}
\label{EMT_ANO_fig}
\end{figure}

FIG.\ref{EMT_ANO_fig} shows $T^{\rho \rho},$ and $T^{\varphi \varphi}$, and $T^{zz}$ for various GL parameter $\kappa = \frac{1}{5}, 0.565, \frac{1}{\sqrt{2}} , 1$, and $\infty$ with a unit winding number.

\begin{figure*}[t]
\centering
\includegraphics[width=0.65\textwidth]{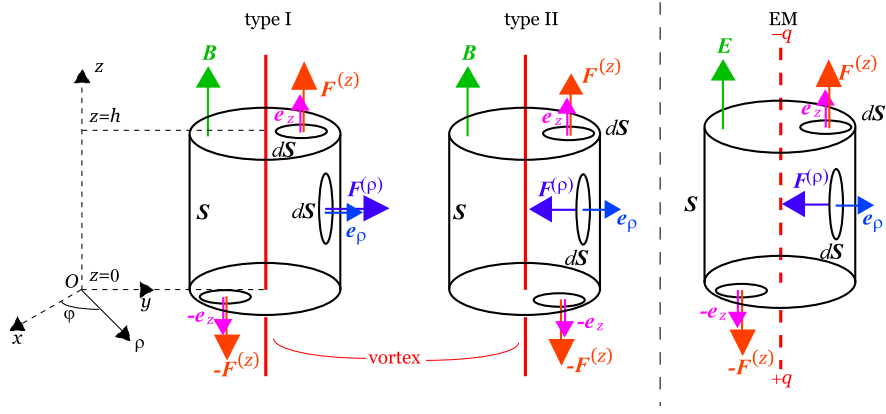}
\caption{(Left and Mid panels) The Maxwell stress force acting on the flux tube originating from the ANO vortex configuration. 
(Right panel) The Maxwell stress force in the electromagnetism.
Here, $h$ represents a height of the cylinder.}
\label{force_EM}
\end{figure*}
\begin{figure*}[t]
\centering
\includegraphics[width=0.65\textwidth]{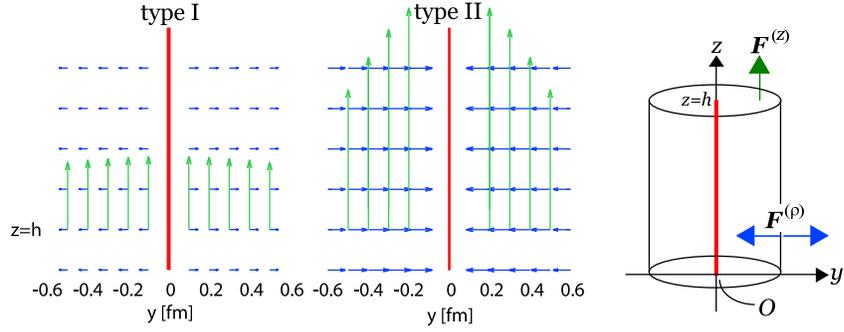}
\caption{The  distribution of the stress forces $\bm{F}^{(\rho)}$ and $\bm{F}^{(z)}$ on the $x = 0$ plane for (Left panel) type I ($\kappa = 0.565$) and (Right panel) type I\hspace{-.1em}I ($\kappa = 1$).
We have illustrated the stress forces around the cross section of the flux tube at $z = h$.
The lengths of the arrows stand for the relative magnitude of the stress forces.
The red line stands for the vortex.}
\label{force_xz}
\end{figure*}

One finds that $T^{\rho \rho}$ is always positive in type I, while always negative in type I\hspace{-.1em}I:
\begin{align}
&T^{\rho \rho} (R) > 0 \ \ \ \left(\kappa < \frac{1}{\sqrt{2}} \right) , \label{Trr_1}
\end{align}
\begin{align}
&T^{\rho \rho} (R) < 0 \ \ \ \left( \kappa > \frac{1}{\sqrt{2}} \right) 
.
\label{Trr_2}
\end{align}
At the boundary between them, i.e., the BPS limit $\kappa = \frac{1}{\sqrt{2}}$, $T^{\rho \rho}$ is identically  zero:
\begin{equation}
T^{\rho \rho} (R) \equiv 0  \ \ \ \left( \kappa = \frac{1}{\sqrt{2}} \right)
.
\end{equation}

It should be noticed that the components $T^{\rho\rho}$ and $T^{\varphi \varphi}$ are not independent, since the conservation law of the Noether current $\partial^{\mu} T_{\mu \nu} = 0$ leads to
\begin{equation}
T^{\varphi \varphi} (R) = \frac{d}{d R} \biggl[ R T^{\rho \rho} (R) \biggr]
.
\end{equation}
This yields that the sign of $T^{\varphi \varphi} (R)$ flips and hence there is a critical value $R = R_{\ast}$ where $T^{\varphi \varphi} (R_{\ast}) = 0$.
See the middle panel of FIG.\ref{EMT_ANO_fig}.

Next, we investigate the force acting on the area element of the flux tube.
By using the Maxwell stress tensor, the stress force $\bm{F}$ acting on the infinitesimal area element $d \bm{S}$ is given by
\begin{equation}
\bm{F} = T \cdot d \bm{S} = T \cdot \bm{n} \Delta S
,
\end{equation}
where $\bm{n}$ is a normal vector perpendicular to the area element $d S$, and $\Delta S$ stands for the area of $d S$.
See FIG.\ref{force_EM}.
The left and mid panels show the situations for the ANO vortex, while the right panel shows the corresponding situation in the electromagnetism, where a pair of electric charges $\pm q$ is located at $\mp \infty$ on the $z$-axis.

If we choose $\bm{n}$ to be equal to the normal vector pointing the $\rho$-direction, i.e., $\bm{n} = \bm{e}_{\rho}$, the corresponding stress force $\bm{F}^{(\rho)}$ reads
\begin{equation}
\bm{F}^{(\rho)} = T^{\rho \rho} \Delta S \bm{e}_{\rho}
.
\end{equation}
Since $T^{\rho \rho}$ obeys (\ref{Trr_1}) and (\ref{Trr_2}), we observe that $\bm{F}^{(\rho)} \cdot \bm{e}_{\rho} = T^{\rho \rho} \Delta S$ is always positive in type I, while always negative in type I\hspace{-.1em}I. 
Therefore, we find that $\bm{F}^{(\rho)}$ represents the attractive force for type I, while the repulsive force for type I\hspace{-.1em}I.

If we choose $\bm{n}$ as the unit vector for the $\varphi$-direction, $\bm{n} = \bm{e}_{\varphi}$, the corresponding stress force $\bm{F}^{(\varphi)}$ is written as
\begin{equation}
\bm{F}^{(\varphi)} = T^{\varphi \varphi} \Delta S \bm{e}_{\varphi}
.
\end{equation}
The sign of $\bm{F}^{(\varphi)} \cdot \bm{e}_{\varphi} = T^{\varphi \varphi} \Delta S$ changes, since the sign of $T^{\varphi \varphi}$ flips at some critical value $R = R_{\ast}$.
This feature could be an artifact due to the infinite length of the ANO vortex and should be investigated in a more realistic situation.

The other choice of $\bm{n}$ is to be parallel to the ANO vortex, i.e., $\bm{n} = \bm{e}_{z}$.
The corresponding stress force $\bm{F}^{(z)}$ can be written as
\begin{align}
\bm{F}^{(z)} = T^{z z} \Delta S \bm{e}_{z} , \ \ \ 
\bm{F}^{(z)} \cdot \bm{e}_{z} = T^{zz} \Delta S > 0
.
\end{align}

FIG.\ref{force_xz} shows the distribution of the stress forces $\bm{F}^{(\rho)}$ and $\bm{F}^{(z)}$ in $y-z$ plane.
Therefore, $\bm{F}^{(z)}$ represents the {attractive force}.
Since $T^{zz}$ is always positive $T^{zz} > 0$ due to (\ref{Tzz}), $\bm{F}^{(z)}$ points the same direction regardless of the value of the GL parameter $\kappa$.

It should be noted that the situation of the type I\hspace{-.1em}I superconductor is similar to the electromagnetism, see the mid and right panels of FIG.\ref{force_EM}.

\begin{figure}[t]
\centering
\includegraphics[width=0.45\textwidth]{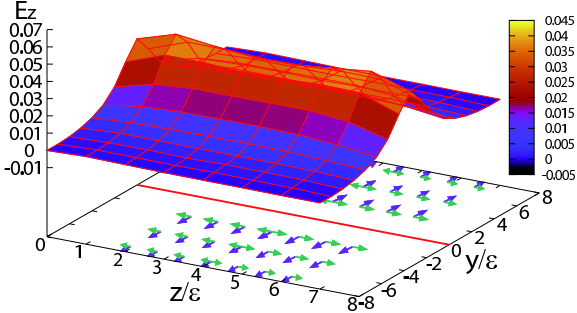}
\caption{The chromoelectric flux obtained in \cite{Kato-Kondo-Shibata} and the distribution of the Maxwell stress forces $\bm{F}^{(\rho)}$ and $\bm{F}^{(z)}$ for the fitted value of the GL parameter $\kappa = 0.565$.
We have taken the height of the cylinder as $h = 8 \epsilon$ to correspond to the distance between the static sources.
The red line (the thick line in the $y-z$ plane) stands for the ANO vortex.
}
\label{distribution}
\end{figure}

Using the parameters obtained by fitting to the ANO vortex, we can reproduce the distribution of the Maxwell stress force around the flux tube, which is shown in FIG.\ref{distribution}.
This result indeed supports the type I dual superconductor for quark confinement.

Our analysis on the Maxwell stress tensor around an ANO vortex agrees with the result obtained by the preceding work \cite{EMT,EMT2-1,EMT2-2,EMT2-3}.

\section{Conclusion}

In this paper, we have studied the type of dual superconductivity for the $SU(2)$ Yang--Mills theory based on the new method:
\begin{itemize}
\item We have introduced the restricted field $V$ to extract the dominant mode for quark confinement and define the induced magnetic current in a gauge-invariant way.

\item We have solved the field equations of the ANO vortex in the $U(1)$ gauge-scalar model numerically without any approximations.
The previous method is based on the Clem ansatz which assumes an analytic form of the complex scalar field without solving the field equations.

\item We have used the resulting magnetic field and the electric current to fit respectively the chromoelectric flux tube and the induced magnetic current obtained by lattice simulations.
In the previous method, only the regression for the chromoelectric flux tube was considered.
\end{itemize}

We have reconfirmed that the vacuum of the $SU(2)$ Yang--Mills theory is of  {\it type I} as a dual superconductor with the GL parameter $\kappa = 0.565 \pm 0.053$ by using the new method.

This result obtained by the new method should be compared with the previous method based on the Clem ansatz:
\begin{itemize}
\item We found that the result of type I agrees with \cite{Kato-Kondo-Shibata} reproduced and supplemented by (\ref{Clem1})--(\ref{Clem_4}).
In the new method, we determined the GL parameter with good accuracy.
\item We have investigated the sensitivity for the fitting range.
We found that the inclusion of the short range modifies the value of the GL parameter $\kappa$ to a smaller one under the Clem ansatz.
For the new method, on the other hand, we found that the inclusion or exclusion of the short range does not effect the GL parameter $\kappa$.
\item We also found that the new method proposed in this paper improves the accuracy of the fitting as seen from the error of the GL parameter, or the mean of squared residuals in both methods.
Therefore, the inclusion of the regression of the magnetic current is important.
\end{itemize}

Moreover, we have obtained the distribution of the Maxwell stress force around the flux tube by using the obtained GL parameter.
It was observed that there exists the attractive force among the chromoelectric flux tubes, which also supports the type I dual superconductor.

\section*{Acknowledgement}
The authors would like to thank Hideo Suganuma for valuable discussions, especially suggestions on error estimations.
They would like to express sincere thanks to Ryosuke Yanagihara, Takumi Iritani, Masakiyo Kitazawa, and Tetsuo Hatsuda for very helpful and illuminating discussions on the Maxwell stress tensor in the early stage of their investigations, on which a part of the result presented in section V is based.
This work was supported by Grant-in-Aid for Scientific Research, JSPS KAKENHI Grant Number (C) No.15K05042 and No.19K03840.
S.N. thanks Nakamura Sekizen-kai for a scholarship.

\appendix
\section{Restricted field variable in the new formulation}

In this paper, we have used the new formulation \cite{KSSMKI08,SKS10} of the lattice Yang--Mills theory: the gauge field variable $U_{x, \mu}$ is decomposed into $V_{x, \mu}$ and $X_{x, \mu}$, $U_{x, \mu} = X_{x, \mu} V_{x, \mu}$.
Here $V_{x, \mu} \in SU (2)$ is called the {\it restricted link variable} which has the same transformation law as the original link variable $U$ under the gauge transformation, while $X_{x, \mu} \in SU(2)$ is a remaining part called the {\it remaining site variable} which transforms in an adjoint way under the gauge transformation. 

In this decomposition we introduce the so-called {\it color direction field} $\bm{n}_{x}$ which is a  Lie-algebra valued field written using the Pauli matrices $\sigma_A$ ($A=1,2,3$) as 
\begin{align} 
 \bm{n}_{x} = n_{x}^A \sigma_A   ,
\end{align}
and is subject to the condition of unit length 
\begin{align} 
\bm{n}_{x} \cdot \bm{n}_{x} :=\bm{n}_{x}^A  \bm{n}_{x}^A =\frac{1}{2} {\rm tr}(\bm{n}_{x}   \bm{n}_{x})=1 .
\end{align}
Therefore, the color direction field $\bm{n}_{x}$ does not cover the whole $SU(2)$ and takes the value only in the coset $G/H=SU(2)/U(1)$. 
An expression exhibiting manifestly this fact is given using $\Omega_{x} \in SU(2)$ as  
\begin{align} 
 \bm{n}_{x} = \Omega_{x} \sigma_3 \Omega_{x}^\dagger \in Lie(SU(2)/U(1))) ,
\end{align}
which is supposed to transform in an adjoint way under the gauge transformation. 
The color direction field $\bm{n}_{x}$ is obtained by solving the reduction condition for a given set of the original link variables $U_{x,\mu}$. 
Therefore, the color direction field $\bm{n}_{x}$ is understood as a functional of the original link variables $U_{x, \mu}$. 
See, e.g., \cite{PhysRept} for a review.

First, the expression (\ref{sol}) for the restricted link variable $V_{x,\mu}$ is obtained \cite{IKKMSS06} by solving the defining equations:  
\begin{align}
\bm{n}_{x} V_{x , \mu} = V_{x, \mu} \bm{n}_{x + \mu}
,
\label{def-L1}
\\
 {\rm tr}({\bf n}_{x} X_{x,\mu}  ) 
  = 0 .
  \label{def-L2}
\end{align} 
By solving the  defining equations (\ref{def-L1}) and (\ref{def-L2}), indeed, 
the link variable $V_{x,\mu}$ is determined  up to an overall normalization constant 
in terms of the site variable $\bm{n}_{x}$ and the original link variable 
$U_{x,\mu}$ as $\tilde{V}_{x,\mu}$: 
\begin{align} 
  \tilde{V}_{x,\mu} 
  := U_{x,\mu} +   \bm{n}_{x} U_{x,\mu} \bm{n}_{x+\mu} ,
\end{align}
However, the defining equation (\ref{def-L1}) is linear in $V_{x,\mu}$. Therefore, the normalization of $V_{x,\mu}$ cannot be determined by the defining equation alone.  Consequently, unitarity is not guaranteed for the general solution of the defining equation and hence a unitarity condition must be imposed afterwards. 
Fortunately, this issue is easily solved at least for the $SU(2)$ group, since the speciality condition $\det V_{x,\mu}=1$ determines the normalization. 
Then the special unitary link variable $V_{x,\mu}$ is obtained after the normalization of $\tilde{V}_{x,\mu}$ as  
\begin{align} 
V_{x,\mu} = 
 \tilde{V}_{x,\mu}/\sqrt{\frac{1}{2}{\rm tr} [\tilde{V}_{x,\mu}^{\dagger}\tilde{V}_{x,\mu}]} .
\end{align}
It is also shown \cite{IKKMSS06} that the naive continuum limit $\epsilon \rightarrow 0$ 
of the link variable $V_{x,\mu} = \exp (-i\epsilon g \mathscr{V}_\mu(x))$ reduces to the continuum expression:
\begin{align} 
\mathscr{V}_{\mu}(x) = (n^A(x)A_{\mu}^A(x))\bm{n}(x)
-ig^{-1} [\partial_{\mu}\bm{n}(x) ,  \bm{n}(x) ] ,
\label{cfn-conti-7} 
\end{align}
which  agrees with the expression of the restricted field in the Cho-Duan-Ge decomposition  in the continuum \cite{Cho80-1,Cho80-2,DG79}. 
This is indeed the case for the remaining variable $X_{x,\mu} = \exp (-i\epsilon g  \mathscr{X}_\mu(x))$.
The $SU(N)$ group is treated in \cite{KSSMKI08,SKS10,PhysRept}.

Second, we focus on the fact that the color direction field $\bm{n}_{x}$ is covariantly constant under the restricted link variable $V_{x, \mu}$ by construction (\ref{def-L1}).
The meaning of ``covariantly constant'' is that one can perform the parallel transport of a vector along a path from point $x$ to point $y$, so that the result is independent of the path chosen.
In particular, parallel transport along a closed loop should leave the vector unchanged.
Suppose we parallel transport $\bm{n}_{x}$ from a point $x$ to a point $x + \mu + \nu$ via two different paths: (I) from $x$ to $x + \mu$ to $x + \mu + \nu$, and (II) from $x$ to $x + \nu$ to $x + \mu + \nu$.
We quickly see  from  (\ref{def-L1}) that
\begin{align}
\bm{n}_{x+ \mu + \nu} = & V_{x+\mu , \nu}^{\dagger} V_{x, \mu}^{\dagger} \bm{n}_{x} V_{x , \mu} V_{x+\mu , \nu} \ \ \ \ \ \ ({\rm path \ 1}) 
\nonumber\\
= & V_{x+ \nu, \mu}^{\dagger} V_{x , \nu}^{\dagger} \bm{n}_{x} V_{x , \nu} V_{x+ \nu , \mu} \ \ \ \ \ \ ({\rm path \ 2}) .
\end{align}
Equating the right-hand side of the first and second lines, and using the unitarity of the $V$'s,
we find that the color direction field $\bm{n}_{x}$ at site $x$ satisfies the relation
\begin{equation}
 \bm{n}_{x} = V_{P_x}^\dagger \bm{n}_{x } V_{P_x} 
 ,
\label{nP=Pn}
\end{equation}
where $V_{P_x}$ is the plaquette variable at $x$, namely, the product of link variables starting at $x$ along the plaquette $P$. 
The relation (\ref{nP=Pn}) is equivalent to 
\begin{equation}
 [V_{P_x},  \bm{n}_{x} ] = 0  
,
\label{Vn=0}
\end{equation}
due to the unitarity of $V_{P_x}$, $V_{P_x}V_{P_x}^\dagger=1$. 
This consideration can be generalized: 
Let $C_{x}$ be any contour on the lattice beginning and ending at site $x$, and let $V_{C_{x}}$ be the holonomy equal to the product of $V$ link variables around the loop $C_{x}$.
It is not hard to see that (\ref{def-L1}) implies
\begin{equation}
\bm{n}_{x} = V_{C_{x}}^{\dagger} \bm{n}_{x} V_{C_{x}} 
\Leftrightarrow 
[V_{C_{x}},  \bm{n}_{x} ] = 0
.
\label{VCn}
\end{equation}
This equation holds at every site $x$ for every possible contour $C_{x}$.
A naive inspection of these equations would yield the view that the equation (\ref{VCn}) following from (\ref{nP=Pn}) or (\ref{Vn=0}) is possible only when the $V$ field is a pure gauge with vanishing field strength.
However, this is not true, because  
 an $SU(2)$ group element $V \in SU(2)$ obeys the identity, 
\begin{equation}
V = \frac{\mathrm{tr}( V)}{\mathrm{tr}(\bm{1})} \bm{1} + 2 \mathrm{tr} \left( V \bm{n}  \right)  \bm{n} 
.
\label{Vp}
\end{equation}
See e.g., \cite{SKS10} for a proof of this identity. 
Indeed, this form of $V$, a linear combination of a unit matrix $\bm{1}$ and the color direction field $\bm{n} $ is consistent with (\ref{Vn=0}) and (\ref{nP=Pn}). 

Recall that we have already given the explicit expression for the restricted link variable $V_{x,\mu}$ as (\ref{sol}):
\begin{align} 
V_{x,\mu} = 
 \tilde{V}_{x,\mu}/\sqrt{\frac{1}{2}{\rm tr} [\tilde{V}_{x,\mu}^{\dagger}\tilde{V}_{x,\mu}]} ,
\nonumber\\
  \tilde{V}_{x,\mu} 
  := U_{x,\mu} +   \bm{n}_{x} U_{x,\mu} \bm{n}_{x+\mu} .
  \label{sol3}
\end{align}
This result shows that the product $\prod_{<x,\mu> \in C} V_{x,\mu}$ of $V_{x,\mu}$ along the loop $C$   does not agree with the pure gauge. 
However, it should be remarked that this conclusion is due to the special property of the color direction field $\bm{n}_{x}$ which we have adopted in this construction as mentioned above: The color direction field $\bm{n}_{x}$ takes the value only in the coset $G/H=SU(2)/U(1)$, 
\begin{align} 
 \bm{n}_{x} \in Lie(SU(2)/U(1))) .
\end{align} 
In contrast, if we start from a field $\hat{\phi}_{x}$ taking the value in the whole $SU(2)$, 
\begin{align} 
 \hat{\phi}_{x} \in Lie(SU(2)) ,
\end{align} 
and impose the covariant-constant condition 
\begin{align}
\hat{\phi}_{x} V_{x , \mu} = V_{x, \mu} \hat{\phi}_{x + \mu}
,
\end{align} 
then the restricted variable just agrees with the pure gauge 
\begin{align}
 V_{x , \mu} =  \text{pure gauge} .
\end{align} 
This case was indeed shown in the continuum formulation \cite{Kondo18}. 

The above remark is also understood in the continuum version of the new formulation \cite{PhysRept} which agrees with the naive continuum limit obtained by taking the limit of vanishing lattice spacing $\epsilon \to 0$ in the lattice version.  
The restricted plaquette variable is expanded as
\begin{align}
V_{P_x} = & \exp \left( -ig\epsilon^2 \mathscr{F}_{\mu\nu} [\mathscr{V}] (P_x) \right) \nonumber\\
=& \bm{1} -ig\epsilon^2 \mathscr{F}_{\mu\nu} [\mathscr{V}] (x) + \mathscr{O}(\epsilon^4) 
.
\label{Vpexp}
\end{align}
The continuum field strength $\mathscr{F}_{\mu\nu} [\mathscr{V}] (x)$ of the restricted field $V_\mu(x)$ is proportional to the color direction field $\bm{n}(x)$ by construction:
\begin{align}
 \mathscr{F}_{\mu\nu} [\mathscr{V}] (x) =  f_{\mu\nu}(x) \bm{n}(x)  
.
\label{FFn}
\end{align}
This implies that the gauge transformation  $\mathscr{F}_{\mu\nu} [\mathscr{V}] (x) \to \Omega(x) \mathscr{F}_{\mu\nu} [\mathscr{V}] (x) \Omega(x)^\dagger$ of the field strength $\mathscr{F}_{\mu\nu} [\mathscr{V}] (x)$ is carried by the color direction field $\bm{n}(x)$ which transforms in the adjoint way $\bm{n}(x) \to \Omega(x) \bm{n}(x) \Omega(x)^\dagger$ under the gauge transformation  $\Omega(x) \in SU(2)$ so that the field strength $f_{\mu\nu}(x)$ defined using the color direction field $\bm{n}(x)$ with a unit length 
$\bm{n}(x) \cdot \bm{n}(x)=1$ by
\begin{align}
f_{\mu\nu}(x) = \bm{n}(x) \cdot \mathscr{F}_{\mu\nu} [\mathscr{V}] (x) := 2{\rm tr}(\bm{n}(x) \mathscr{F}_{\mu\nu} [\mathscr{V}] (x) )
\end{align}
is invariant under the gauge transformation. 
Note that (\ref{Vpexp}) and (\ref{FFn}) are consistent with (\ref{Vp}), since in the continuum limit, $2 \mathrm{tr} \left(  \bm{n}_{x} V_{P} \right)$ reduces to the gauge-invariant field strength $\bm{n}(x) \cdot \mathscr{F}_{\mu \nu} [\mathscr{V}](x)$  
\begin{align}
 2 \mathrm{tr} \left(  \bm{n}_{x} V_{P} \right)  
 =& 2 \mathrm{tr} \left(  \bm{n}_{x} \right)   -ig\epsilon^2 2 \mathrm{tr} \left(  \bm{n}_{x}  \mathscr{F}_{\mu\nu}[\mathscr{V}](P) \right)  + \mathscr{O}(\epsilon^4) 
 \nonumber\\
=& -ig\epsilon^2 \bm{n}(x) \cdot \mathscr{F}_{\mu \nu} [\mathscr{V}](x) + \mathscr{O}(\epsilon^4) 
.
\end{align}
Therefore, (\ref{Vn=0}) does not mean that the $V$ field is a pure gauge with vanishing field strength $\mathscr{F}_{\mu\nu} [\mathscr{V}]$. 
Thus we have the non-trivial gauge-invariant field strength $f_{\mu\nu}(x)$ which is used to measure the chromoelectrix flux. 
This feature is more clearly seen using the non-Abelian Stokes theorem for the Wilson loop operator, see e.g., \cite{PhysRept}. 

Finally, in oder to see the physical meaning of the restricted operator $\rho [V]$ and the independence from the Schwinger lines to be inserted, we cast it into another form.  
The covariant constantness (\ref{def-L1}) of the color direction field $\bm{n}_{x}$ under the restricted link variable $V_{x , \mu}$ yields another expression for $\rho [V]$.  
In fact, we apply (\ref{Vp}) to $V_{P}$ in (\ref{lattice_operator_V}) and use the property (\ref{def-L1}) to parallel transport the color direction field $\bm{n}_{x}$ on the plaquette $P$ to $\bm{n}_{z}$ at the point $z$ on the line connecting a pair of quark and antiquark, see Fig.~\ref{lattice_result1}: 
\begin{align}
\rho [V] = & \frac{\left\langle 2 \mathrm{tr} \left(  V_{P} \bm{n}_{x}  \right) \mathrm{tr} \left(  W [V] L_{V} \bm{n}_{x} L_{V}^{\dagger} \right) \right\rangle}{\langle \mathrm{tr} (W [V]) \rangle}
\nonumber\\
=& \frac{\left\langle 2 \mathrm{tr} \left(  V_{P} \bm{n}_{x}  \right) \mathrm{tr} \left(  W [V] \bm{n}_{z} L_{V} L_{V}^{\dagger} \right) \right\rangle}{\langle \mathrm{tr} (W [V]) \rangle}
\nonumber\\
=& \frac{\left\langle 2 \mathrm{tr} \left(  V_{P} \bm{n}_{x}  \right) \mathrm{tr} \left(  W [V] \bm{n}_{z} \right) \right\rangle}{\langle \mathrm{tr} (W [V]) \rangle} ,
\end{align}
where we have used the unitarity of $L_{V}$, $L_{V}L_{V}^\dagger=1$ in the last step. 

Thus we find that $\rho [V]$ does not depend on both the choice of the Schwinger lines $L, L^\dagger$ and the position $z$ at which the color direction field is inserted. 
This is not the case for $\rho [U]$ constructed from the original gauge variable $U$.  
These facts demonstrate advantages of using quantities like $\rho [V]$ constructed from the restricted variable based on the new formulation. 


\end{document}